\documentclass[%
 reprint,
superscriptaddress,
 amsmath,amssymb,
 aps,
]{revtex4-2}

\usepackage[T1]{fontenc}
\usepackage{Journal_names}
\usepackage{xcolor}
\usepackage{graphicx}
\usepackage{dcolumn}
\usepackage{bm}
\usepackage{amsmath}
\usepackage{xspace}
\usepackage{tikz}
\usetikzlibrary{fadings, shapes.geometric}
\usepackage{tabularx}
\usepackage{booktabs}
\usepackage{array}
\usepackage{fleqn}

\usepackage{ragged2e}
\usepackage{subcaption}
\makeatletter
\long\def\@makecaption#1#2{%
  \vskip\abovecaptionskip
  \centering 
  \parbox{\linewidth}{\justifying\small #1. #2\par}%
  \vskip\belowcaptionskip}
\makeatother

\newcolumntype{P}[1]{>{\centering\arraybackslash}p{#1}}
\usepackage[colorlinks=true, linkcolor=blue, citecolor=blue, urlcolor=blue]{hyperref}
\usepackage[mathlines]{lineno}

\newcommand{\multistar}{{\tt multistar}\xspace}
\newcommand{\phaseflow}{{\tt PHASEFLOW}\xspace}
\newcommand{\agama}{{\tt AGAMA}\xspace}
\newcommand{\pc}{\ensuremath{\mathrm{\ pc}}}

\newcommand{\dd}{\mathrm{d}}
\newcommand{\Gyr}{\mathrm{Gyr}}
\newcommand{\rin}{\ensuremath{r_{\rm in}}}
\newcommand{\mSMBH}{\ensuremath{M_{\rm BH}}}
\newcommand{\mUR}{\ensuremath{M_{\rm unrelaxed}}}
\newcommand{\RSMBH}{\ensuremath{R_{\rm BH}}}
\newcommand{\mEMRI}{\ensuremath{M_{*}}}
\newcommand{\mDM}{\ensuremath{m_{\rm DM}}}
\newcommand{\rEMRI}{\ensuremath{r_{\rm E}}}
\newcommand{\rp}{\ensuremath{r_{\rm p}}}
\newcommand{\rpEMRI}{\ensuremath{\mathfrak{r}_{\rm p}}}
\newcommand{\rpLISA}{\ensuremath{\mathfrak{r}_{\rm p, LISA}}}
\newcommand{\rpEMRIi}{\ensuremath{\mathfrak{r}_{{\rm p},i}}}
\newcommand{\rpEMRIz}{\ensuremath{\mathfrak{r}_{\rm p, 0}}}
\newcommand{\raEMRI}{\ensuremath{\mathfrak{r}_{\rm a}}}
\newcommand{\rpDM}{\ensuremath{\mathfrak{R}_{\rm p}}}
\newcommand{\raDM}{\ensuremath{\mathfrak{R}_{\rm a}}}
\newcommand{\aEMRI}{\ensuremath{\mathfrak{a}}}
\newcommand{\eEMRI}{\ensuremath{\mathfrak{e}}}
\newcommand{\eEMRIi}{\ensuremath{\mathfrak{e}_i}}
\newcommand{\eEMRIz}{\ensuremath{\mathfrak{e}_0}}
\newcommand{\eDM}{\ensuremath{\mathcal{E}}}

\newcommand{\Msolar}{\ensuremath{\mathrm{M}_{\odot}}}
\newcommand{\Rsolar}{\ensuremath{\mathrm{R}_{\odot}}}
\newcommand{\rout}{\ensuremath{r_{\rm out}}}
\newcommand{\V}[1]{\mathbf{#1}}

\definecolor{cornellGreen}{HTML}{6EB43F}

\begin{document}


\title{The Depletion of Collisionless Dark Matter Spikes}

\author{Charlie Sharpe}
\email{charles.sharpe@physics.ox.ac.uk}
\affiliation{Department of Physics, University of Oxford, Parks Road, Oxford OX1 3PU, United Kingdom}

\author{Yonadav Barry Ginat}
\email{yb.ginat@physics.ox.ac.uk}
\affiliation{Department of Physics, University of Oxford, Parks Road, Oxford OX1 3PU, United Kingdom}

\author{Thomas F.~M.~Spieksma}
\email{thomas.spieksma@physics.ox.ac.uk}
\affiliation{Department of Physics, University of Oxford, Parks Road, Oxford OX1 3PU, United Kingdom}

\author{Alexander Heger}
\email{alexander.heger@monash.edu}
\affiliation{School of Physics and Astronomy, Monash University, Melbourne, Victoria 3800, Australia}

\author{Bence Kocsis}
\email{bence.kocsis@physics.ox.ac.uk}
\affiliation{Department of Physics, University of Oxford, Parks Road, Oxford OX1 3PU, United Kingdom}
\affiliation{St.~Hugh's College, University of Oxford, St.~Margaret's Road, Oxford OX2 6LE, United Kingdom}
\affiliation{Institute of Physics, Eötvös University, Pázmány P.~s.~1/A, Budapest 1117, Hungary}
\date{\today}

\begin{abstract}
Dense concentrations of dark matter surrounding black holes provide a compelling opportunity to probe the nature of dark matter. In the classic Gondolo--Silk model, the adiabatic growth of a massive black hole in a dark matter cusp produces a steep density spike ($\rho \propto r^{-7/3}$), potentially inducing measurable gravitational-wave dephasings in intermediate and extreme mass-ratio inspirals (IMRIs/EMRIs). We challenge this paradigm by considering a collisionless dark matter spike embedded in a realistic nuclear star cluster. Using 1D orbit-averaged Fokker--Planck simulations of isotropic nuclear clusters, we show that mass segregation in a multi-mass stellar cusp accelerates relaxation relative to single-mass models, thereby driving the dark matter to the lower density $r^{-3/2}$ Bahcall--Wolf profile within $\lesssim 1 \Gyr$. In the inner regions, where the Fokker--Planck description breaks down, we model strong triple interactions between dark matter particles and EMRIs using post-Newtonian 3-body simulations. We show that EMRIs eject dark matter particles via gravitational slingshots, depleting the inner spike over a few Gyr. Because EMRI number densities are too low to drive two-body relaxation, and replenishment by DM self-relaxation is negligible on the time-scales considered, this depletion is irreversible. While the extent of EMRI-induced DM depletion depends on the EMRI rate and mass, we find reductions in densities by several orders of magnitude. As a result, the dark-matter-induced dephasings for EMRIs may fall below the LISA detectability threshold for massive black holes at $z = 3$ (2.14 Gyr) with masses $\lesssim 10^{5}\,\Msolar$ (for a low $\mathcal{O}(10) \, \mathrm{Gyr}^{-1}$ EMRI rate), extending to $\lesssim 10^6\,\Msolar$ for more realistic rates of $\mathcal{O}(100 - 300) \, \Gyr^{-1}$. Our findings substantially reduce the parameter space over which massive black holes can host detectable collisionless dark matter spikes.

\end{abstract}

\maketitle


\section{Introduction}
\label{sec:Introduction}


%
Black holes (BHs) may be surrounded by high densities of dark matter (DM) which, if present, could produce distinctive observational signatures. Understanding how such over-densities form and evolve is hence crucial in the search for DM.
In a seminal paper, \citet{Peebles_1972} showed that when an MBH grows adiabatically within a spherical, isothermal stellar core (on a time-scale longer than the dynamical time), adiabatic invariance draws stars into the deepening gravitational potential well, thereby forming a density cusp with $\rho(r) \propto r^{-3/2}$. This was later confirmed numerically by \citet{Young_1980}. These insights, together with early $\gamma$-ray flux observations from the \textit{COS-B} satellite \citep{Blitz_1985, Silk_1987}, motivated the first investigations into whether analogous processes could produce large DM densities around massive objects. \citet{Ipser_1987} were the first to explore this possibility, studying how the growth of the Milky Way’s central spheroid could enhance DM densities and potentially yield detectable $\gamma$-ray signatures. They found, however, that \textit{COS-B} lacked the sensitivity to place meaningful constraints. 
Expanding on this work, \citet{Gondolo_1999} considered a more specialised scenario: the adiabatic growth of an MBH seed embedded in a pre-existing DM cusp. For an initial power-law profile $\rho(r) \propto r^{-\gamma_{\rm 0}}$, they showed that the DM density within the sphere of influence (SoI) --- defined as the region where the enclosed stellar mass equals the mass of the central MBH --- steepens into a power-law $\rho \propto r^{-\gamma_{\rm sp}}$, whose exponent is
\begin{align}
    \gamma_{\rm sp} = \frac{9 - 2\gamma_{\rm 0}}{4 - \gamma_{\rm 0}}\,.
\end{align}
For $\gamma_{\rm 0} \in [0,2]$, this implies $\gamma_{\rm sp} \in [2.25,2.5]$, leading to extremely large DM over-densities, termed ``DM spikes''. This sparked an interest in their phenomenological implications, particularly in $\gamma$-ray and neutrino emission from DM annihilation.
Subsequent work, however, cast doubt on the robustness of these idealised spikes. \citet{Ullio_2001} showed that the \citet{Gondolo_1999} scenario relies on restrictive assumptions --- such as an infinitesimal MBH seed located precisely at the centre of the initial DM cusp --- and that more general configurations typically settle to a $\rho(r) \propto r^{-1/2}$ profile, similar to results obtained for stellar systems \citep{Nakano_1999}. In fact, one must also assume the pre-existing NFW profile is fully formed before the MBH's adiabatic growth; otherwise shallow $\sim r^{-3/2}$ DM profiles can result \citep{Le_Delliou_2010}. Further complications arise from hierarchical structure formation:~\citet{Merritt_2002} demonstrated that equal-mass MBH binary mergers can severely disrupt DM spikes, or even destroy them entirely. Additionally, \citet{Gnedin_2004} and \citet{Merritt_2004} showed that two-body relaxation between DM and a single-mass stellar component drives the system towards a generic lower-density $\rho(r) \propto r^{-3/2}$ profile, even if the system is initially far from equilibrium, consistent with a Bahcall--Wolf (BW) cusp \citep{Bahcall_1976, Bahcall_1977}. While \citet{Merritt_2006} argued that significantly scoured spikes can relax back to a BW profile following mergers, they also showed that the density normalisation decreases steadily as DM is heated and redistributed outwards, dropping by a factor of $1/\mathrm{e}$ over $1.2\,T_{\rm relax}$ and by $1/\mathrm{e}^2$ over $4.5\,T_{\rm relax}$, with $T_{\rm relax}$ being the relaxation time-scale. These conclusions were later confirmed with the $N$-body simulations of Ref. \citep{Merritt_2007}.
More recently, attention has shifted towards intermediate-mass black holes (IMBHs), with masses in the range $10^{2}$--$10^{6}\,\Msolar$, as more promising hosts of DM spikes \citep{Bertone_2005b, Eda_2013}. This follows from the fact that major mergers can destroy spikes \citep{Merritt_2002} and that studies of MBH merger rates \citep{Conselice_2006, Fakhouri_2010, Ravi_2015} indicate that the expected number of major mergers (mass ratio $\geq 0.3$) drops below unity only for BHs with masses $\lesssim 10^{6}$--$10^{7}\,\Msolar$. Similar merger rates can be inferred from predicted galactic merger rates \citep{Rodriguez_Gomez_2016, OLeary_2021} and the correlations between the MBH mass and the stellar mass \citep{Kormendy_2013, Reines_2015}. Consequently, unlike super-massive black holes (SMBHs), IMBHs are unlikely to have experienced major mergers during their lifetime, allowing much greater DM densities to survive --- provided the IMBH grew adiabatically from a seed and hence hosts a DM spike.
Most DM spike studies are phenomenologically motivated and focus on one of two observables. The first concerns particle fluxes produced by self-interacting dark matter (SIDM) and self-annihilating dark matter (SADM), with the specific signatures depending on the underlying particle model.  The absence of bright sources of $\gamma$-rays or neutrinos, however, suggests either that DM is not composed of particles that emit them, or DM spikes do not form  \citep{Bertone_2006, Aharonian_2008, Bertone_2009, Freese_2022, Bertone_2024}. With that being said, shallower DM density profiles, close to $\rho \propto r^{-1}$, are not fully excluded, and could potentially account for the excess $\gamma$-ray emission detected by the \textit{Fermi}-LAT satellite \citep{Hooper_2011a, Hooper_2011b, Abazajian_2012, Daylan_2016, Muru_2025}. 
The second approach is to study the imprint of DM spikes on gravitational waves (GWs), in particular the dephasing of signals emitted by intermediate or extreme mass-ratio inspirals (IMRIs/EMRIs) evolving within the central regions of a DM spike. A wide range of dynamical effects can influence an EMRI embedded in a DM spike. In practice, most studies focus on a subset of effects deemed most relevant to a given system, leading to significant variations in model complexity across the literature. 

For example, many works model the effect of dynamical friction \citep{Eda_2013, Eda_2015, Kavanagh_2020, Hannuksela_2020, Coogan_2022, Cole_2022, Ghoshal_2023, Dosopoulou_2024, Mukherjee_2024, Montalvo_2024, Fischer_2024, Kavanagh_2024, Alonso_Alvarez_2024, Karydas_2024, Tiruvaskar_2025}, two-body scattering \citep{Yue_2019a, Acevedo_2025, Speri_2026, Chen_2026}, accretion of DM onto the lighter BH \citep{Yue_2018, Dai_2022, Karydas_2025}, non-zero orbital eccentricity \citep{Yue_2019b, Becker_2022, Chen_2025}, and relativistic dynamics \citep{Sadeghian_2013, Speeney_2022, Mitra_2025, Vicente_2025}. Despite these differences, a common prediction emerges: DM-induced effects can produce substantial GW dephasing relative to vacuum inspirals. For a single five-year observation with the upcoming Laser Interferometer Space Antenna (LISA), the accumulated phase shift is expected to range between $10^2$ and $10^6$ radians; this is well above the LISA detection threshold \citep{Hinderer_2008}.
Despite significant progress in understanding DM spikes and their detectability, most existing studies rely on two simplifying assumptions:~(i) the DM density slope $\gamma$ is treated as a free parameter in the range $1 \leq \gamma \leq 3$;~and (ii) the DM density is normalised by fixing the density at the radius of influence (RoI or $r_{\rm SoI}$) of the MBH to $10^{2-3} \, \Msolar\,\pc^{-3}$. This normalisation is typically obtained either by extrapolating the local DM density near the Sun inwards under the assumption of a Navarro--Frenk--White (NFW) profile \citep{NFW1997,Gondolo_1999}, or by modelling a high-redshift $10^6 \, \Msolar$ NFW halo and using concentration--mass relations to match the resulting density to an NFW density \citep{Eda_2015}. When combined, these assumptions, particularly when ranging over $\gamma \in [1,3]$, lead to a variation of $10$--$15$ orders of magnitude in the DM density within a few Schwarzschild radii of the MBH --- precisely the region probed by EMRIs in the LISA band \citep{Babak_2017}. 

\begin{figure}
    \centering
    \includegraphics[width=0.45\textwidth]{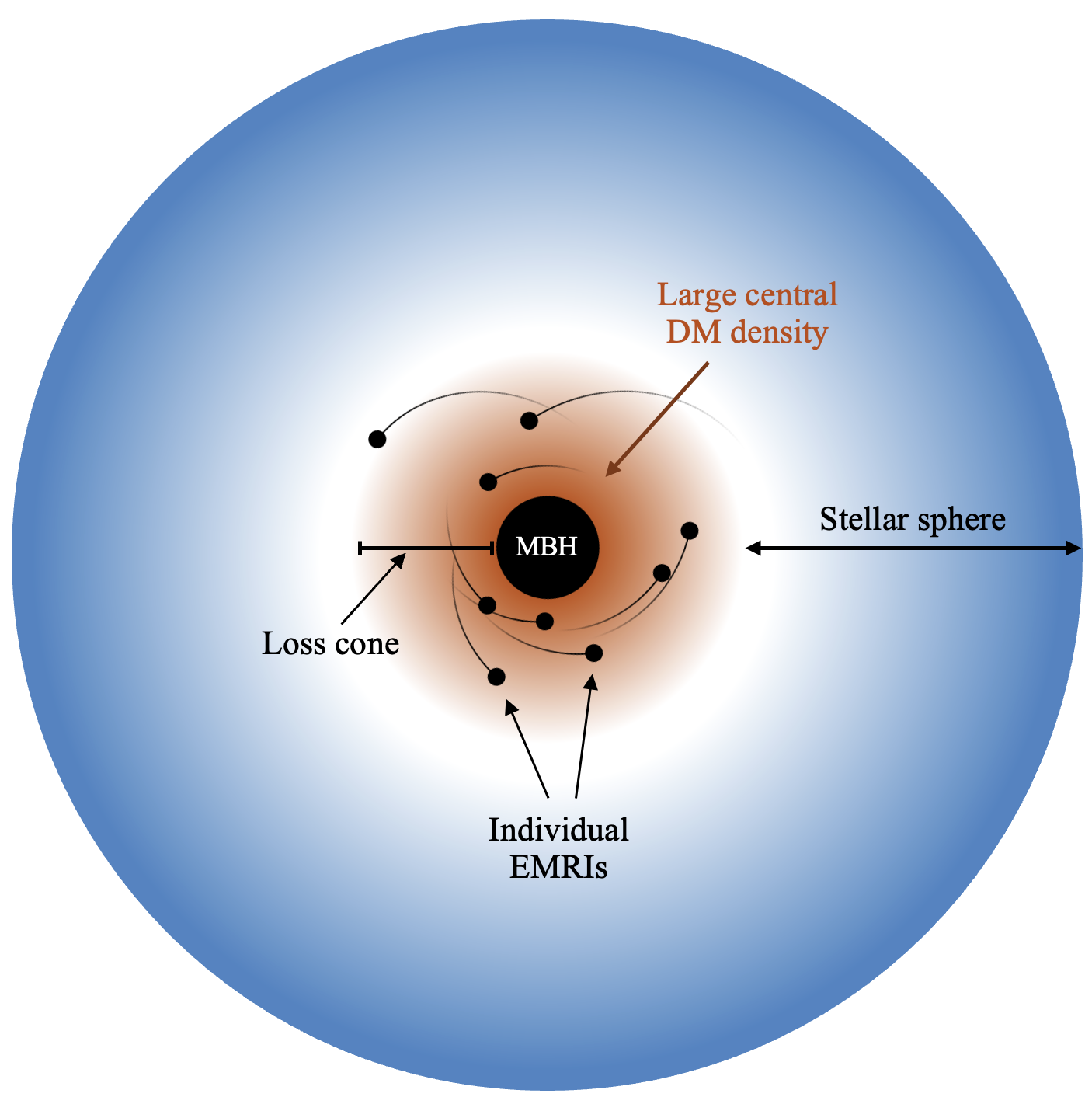}
     \caption{Schematic illustration of our system. The DM spike is shown in brown and the stellar sphere in blue. The full DM distribution (not shown) extends to the outer boundary of the stellar sphere and far beyond it. In reality, the loss cone is not a sphere but a cone in semi-major axis--eccentricity space.}
    \label{fig:System_Config}
\end{figure}

These enormous variations highlight our limited understanding of spike evolution in realistic galactic environments, particularly in the presence of nuclear star clusters (NSCs). Indeed, the commonly adopted choices implicitly decouple the spike from the stellar dynamics of its host system. Addressing these uncertainties is essential for understanding the phenomenological signatures of spikes in realistic astrophysical environments and constitutes the central motivation of this work. We now motivate the two specific concerns we aim to address.
Previous studies have shown that two-body relaxation between DM and stars drives the DM density profile towards $\rho \propto r^{-3/2}$ within a relaxation time. These relaxation processes are typically modelled analytically using orbit-averaged Fokker–Planck (FP) treatments, often benchmarked against $N$-body simulations. However, nearly all FP models of the co-evolution of DM with a stellar component (with the exception of a brief discussion in \citet[][\S5.3]{Mukherjee_2024}) assume that the stellar component is comprised of stars of a single, ``effective'' mass. This assumption is valid only if the relative number density of each stellar species is independent of orbital energy (i.e., distance from the BH; cf. Refs. \citep{Merritt_2004, Gnedin_2004, Merritt_2006, Merritt_2007, Shapiro_2022}). In realistic stellar systems, however, stars and stellar-mass BHs (sBHs) span a broad range of masses, and mass segregation is inevitable. This renders the stellar mass function radius-dependent \citep{Bahcall_1977, Alexander_2009, Preto_2010, Panamarev_2022}, thereby violating the single-mass assumption. Since multi-mass stellar systems relax more efficiently \citep{Perets+2007,OLeary_2009}, we expect them to accelerate the relaxation-induced erosion of the DM spike.
The second, independent limitation concerns the validity of the orbit-averaged FP equation itself. FP theory describes the system’s evolution as driven by the cumulative effect of many small-angle, weak gravitational encounters between stars. This approximation fails in the loss cone, where strong encounters dominate. For example, we find below that for a Sgr A$^*$-like system, this transition occurs at radii of order $10^2 \RSMBH$, where $\RSMBH$ is the MBH's Schwarzschild radius. Inside this region --- and especially at very small radii $\lesssim \mathcal{O}(10) \RSMBH$ --- the evolution of the DM spike is no longer governed by a smooth stellar background, but by the discrete influence of individual compact objects spiralling towards the central BH.
We therefore divide the MBH SoI into two regimes: at large radii, where stellar densities are high enough for FP theory to apply, we define the \textit{stellar sphere}; at smaller radii, dynamics are governed by strong encounters with BHs. This configuration is illustrated in Figure~\ref{fig:System_Config}. The inner, loss-cone region is of particular interest for GW astrophysics, as it is the region where future GW detectors such as LISA~\citep{LISA:2024hlh} become sensitive to the GWs emitted by IMRIs and EMRIs. Meanwhile, the stellar sphere sets the outer boundary conditions for the loss-cone region and can play a crucial role in several contexts: determining EMRI formation rates via dynamical friction at larger radii, establishing conditions for DM mini-spike formation around stellar-mass BHs \citep{Branco_2025}, and influencing the evolution of SMBH binaries, which was argued to affect the nHz stochastic GW background \cite{Ghoshal_2023, Hu_2025, Chen_2025, Shen_2025b}. We do not consider the region outside the SoI of the MBH for two reasons: first, the SoI contains the highest DM densities; second, particle orbits are Keplerian only within the SoI.
In this paper, we pursue two goals: firstly, we investigate how mass segregation within a realistic multi-mass stellar population alters the evolution of a collisionless DM spike around an MBH, compared with the commonly assumed single-effective-mass model. In particular, we study the resulting relaxed density profiles and relaxation time-scales in multi-mass stellar cusps. Secondly, we examine the extent to which repeated EMRI mergers can deplete the DM spike. We assume the initial spike is smooth and isotropic -- a strongly clumped or globally anisotropic DM distribution would require a separate treatment. Key parameters used throughout this work are summarised in Table~\ref{tab:Parameters}.

\begin{table}
    \centering
        \def\sep{0.1ex}
        \begin{tabular}[c]{|c|c|}
            \hline 
            Symbol & Meaning \\ 
            \hline
            $\mSMBH$ & Mass of the central MBH \\ [\sep]
            $\RSMBH$ & $2\mathrm{G} \mSMBH/\mathrm{c}^2$, Schwarzschild radius of the central MBH \\ [\sep]
            $\mEMRI$ & Mass of the sBH EMRI component \\ [\sep]
            $\rpEMRI$ & Periapsis of the EMRI \\ [\sep]
            $\raEMRI$ & Apoapsis of the EMRI \\ [\sep]
            $\aEMRI$ & Semi-major axis of the EMRI \\ [\sep]
            $\eEMRI$ & Eccentricity of the EMRI \\ [\sep]
            $\mDM$ & Mass of an individual DM particle \\ [\sep]
            $\rpDM$ & Periapsis of the DM particle about the MBH \\ [\sep]
            $\raDM$ & Apoapsis of the DM particle about the MBH \\ [\sep]
            $\eDM$ & Eccentricity of the DM particle about the MBH \\ [\sep]
            $\iota$ & Relative inclination between the DM and EMRI orbits \\ [\sep]
            \hline
        \end{tabular}
        \caption{Summary of key variables used in this work.}
    \label{tab:Parameters}
\end{table}

The remainder of this paper is organised as follows: in \S\ref{sec:DM Spikes in Mass Segregated Stellar Cusps}, we use the FP formalism to compare the co-evolution of a DM spike embedded in a multi-mass stellar cusp with that in a single-effective-mass cusp; then, in \S\ref{sec:DM and EMRIs}, we model the gradual evaporation of DM in the loss cone due to slingshot interactions with EMRIs using a three-body framework. We discuss the implications of our results in \S\ref{sec:Discussion} and present our conclusions in \S\ref{sec:Conclusions}. 
%
\section{DM Spikes in Mass-segregated Stellar Cusps}
\label{sec:DM Spikes in Mass Segregated Stellar Cusps}
Let us commence with the co-evolution of stars and DM within the outer region of the SoI --- the stellar sphere. Previous studies have almost exclusively assumed a single stellar mass component, thereby precluding any mass segregation; in reality, multi-mass stellar components are essential for modelling the long-term evolution of a stellar system, as mass segregation alters relaxation times with direct consequences for the resulting DM density profile. We now present a detailed investigation of the evolution of DM spikes in the presence of a multi-mass stellar cusp.

This section is structured as follows. In \S\ref{subsec:FP and SSE}, we describe our FP treatment, its validity and the steady-state equilibria it predicts. In \S\ref{subsec:Time-scales}, we discuss the relevant time-scales of our system before discussing our stellar model and initial density profiles assumed in \S\ref{subsec:Stellar Model and DM initial density}. Finally, we present our results in \S\ref{subsec:Results -- FP}.
\subsection{The Fokker--Planck Equation and Steady-state Equilibrium}
\label{subsec:FP and SSE}

\subsubsection{The Fokker--Planck Equation}
\label{subsubsec:FPE}
The FP equation describes the evolution of the phase-space density of a particle species, $s$, under the assumption that interactions are dominated by distant, weak encounters --- that is, it models the cumulative effect of many repeated small kicks on the phase-space density. Since the collisional time-scale is long compared to the orbital period, the FP equation may be orbit-averaged. Here, we adopt the 1D orbit-averaged, isotropic FP equation in energy space, describing how the orbital energies $E$ of individual stars diffuse in time due to interactions with objects in an isotropic cluster having a thermal eccentricity distribution, $f_e = 2e$ \citep{Merritt_1983}
\begin{align}
    -\frac{\dd q}{\dd E} \frac{\partial f_s}{\partial t}=\frac{\partial}{\partial E}\left[D_{E, s}(E) f_s(E)+D_{EE, s}(E) \frac{\partial f_s}{\partial E}\right]\,, \label{eq:FP}
\end{align}
where $f_s(E,t)$ is the phase-space distribution function of species $s$, $q(E)$ is the phase-space volume enclosed by the orbital energy surface, and $D_{EE,s}(E)$ and $D_{E,s}(E)$ are the energy diffusion and dynamical friction coefficients, respectively. These are given by 
\begin{align}
\begin{aligned}
    q(E) &= \frac{2^{3/2}\pi^3}{3} \mathrm{G}^3 \mSMBH^3 E^{-3/2}\,,\\
    D_{E E, s}(E) &=A_0 \sum_{k} m_k^2\left(\int_E^{\infty} f_k\left(E_k\right) q\left(E_k\right) \mathrm{d} E_k + \right. \\
    & \left.q(E) \int_{-\infty}^E f_k\left(E_k\right) \mathrm{d} E_k\right)\,, \\
    D_{E, s}(E) &=A_0 m_s \sum_{k}  m_k\left(\int_E^{\infty} f_k\left(E_k\right) \frac{\partial q}{\partial E_k} \mathrm{d} E_k\right)\,,\\
    A_0 &= 64 \pi^4 \mathrm{G}^2 \ln \Lambda\,.
\end{aligned} \label{eq:q,D_EE,D_E}
\end{align}
The sums run over all species $k$ that interact gravitationally with species $s$, and $\ln \Lambda$ is the Coulomb logarithm which we have taken to be 15 \citep{Vasiliev_2017}.
In this form, the equation neglects strong encounters, Newtonian apsidal precession, relativistic effects (such as relativistic apsidal precession and detailed loss cone dynamics), and resonant effects including scalar and vector resonant relaxation. However, these omissions are justified for our purposes. While individual strong encounters can generate direct plunges and induce order-unity changes in orbital energies, they have been shown to have negligible impact on the global evolution of stellar cusps \citep{Babak_2017, Rom_2024, Zhong_2023,Amaro_Seoane_2025}. Scalar resonant relaxation (SRR) accelerates the relaxation of eccentricities towards a maximal entropy distribution, that is, towards $f(e) = 2e$ \citep{Rauch_1996}. Since our 1D FP treatment already assumes this equilibrium distribution, SRR does not modify our results and, if anything, supports the consistency of our approach. Moreover, while early studies suggested that SRR could strongly affect loss-cone dynamics \citep{Hopman_2006a}, subsequent work has demonstrated that SRR is quenched by relativistic apsidal precession and hence does not play a major role in the steady-state dynamics of the loss cone \citep{Kocsis_2011, Merritt_2011, Brem_2012, Merritt_2015, Bar_Or_2016, Alexander_2017, Fouvry_2017, Bar_Or_2018, Generozov_2020, Emami_2020}. Vector resonant relaxation (VRR), by definition, only changes the orientation of orbital angular momentum vectors (i.e., relaxes angular momentum directions) and thus affects neither the eccentricities nor the orbital energies.
Independent support for the FP approach comes from direct $N$-body simulations \citep{Kim_2008, Amaro_Seoane_2011, Vasiliev_2017}, which show excellent agreement with FP predictions down to the smallest scales that can be reliably resolved numerically \footnote{Note that \citet{Vasiliev_2017} finds disagreement at $r \lesssim 0.01 r_{\rm SoI}$ and ascribes this discrepancy to resonant relaxation in the $N$-body model. However, their $N$-body model is Newtonian, and we thus do not expect agreement unless relativistic apsidal precession is included.}. Similarly, \citet{Panamarev_2019} model the Galactic Centre using $N$-body simulations and obtain density profiles in close agreement with the BW solution and hence with FP predictions.
A fully self-consistent treatment incorporating strong encounters, relativistic dynamics, SRR, and VRR is beyond the scope of this work. Nevertheless, the considerations above justify neglecting these effects and support the use of the standard FP formalism down to the small radii considered here.
We solve the one-dimensional orbit-averaged FP equation using a finite-element numerical integration scheme implemented in the publicly available code \phaseflow \citep{Vasiliev_2017}, which is part of the \agama package \citep{Vasiliev_2018}. This enables us to quantify the differences in relaxation times between single-mass and multi-mass stellar cusps.

\subsubsection{Steady-state Equilibrium}
\label{subsubsec:SSE}

After many relaxation times, the system approaches a steady-state in which each stellar or particle species settles into a power-law density profile whose exponent depends on its mass. For an isotropic system in a Keplerian potential, one can obtain these slopes using the close relationship between energy-space density and radial density: if $f(E) \propto E^{p}$ then $\rho \propto r^{-p - 3/2}$ \citep[][Eq.~(4.43)]{Binney_2008}.
From Eq.~\eqref{eq:q,D_EE,D_E}, we obtain
\begin{align}
    \frac{D_{E,s} f_s(E)}{D_{EE, s}\frac{\partial f_s}{\partial E}} \sim \frac{m_s \sum_k m_k}{\sum_k m_k^2}\,.
\end{align}
Since $\mDM \ll m_{k}$, this implies the dynamical friction acting on the DM component is negligible, and its evolution is governed purely by diffusion in energy space:
\begin{align}
    - \frac{\partial}{\partial t} \left(\frac{\dd q}{\dd E} f_{\rm DM}\right) = \frac{\partial}{\partial E}\left[D_{E E, {\rm DM}}(E) \frac{\partial f_{\rm DM}}{\partial E}\right]\,.
\end{align}
We identify
\begin{align}
    F_{E,{\rm DM}} := D_{E E, {\rm DM}}(E) \frac{\partial f_{\rm DM}}{\partial E}\,,
\end{align}
as the energy-space flux of DM particles. In steady state, the system approaches a zero-net-energy-flux configuration, $F_{E,{\rm DM}} = 0$ (see \footnote{This condition is violated near the loss cone, where particles are lost to the central MBH.}). Consequently, $\partial f_{\rm DM}/\partial E = 0$, implying $p_{\rm DM} = 0$ and hence $\rho_{\rm DM} = r^{-3/2}$ \citep{Bahcall_1977,Alexander_2009,Linial_Sari2022,Rom_Linial_Sari_2023}. 
The equilibrium profile of the stars, by contrast, depends on whether the system exhibits weak or strong mass-segregation. Below, we summarise analytical arguments from the literature that derive the steady-state density slopes in each regime for a two-stellar-mass-component system \citep[see][for the more general many-mass case]{OLeary_2009,Keshet_2009,Vasiliev_2017,Linial_Sari2022}.
In the case of weak mass segregation, the heavy stars are sufficiently numerous that they predominantly scatter off one another, rather than the lighter stars. Assuming these heavy stars have density profile $\rho_{\rm H} \propto r^{-\alpha}$, then within the SoI --- where the velocity dispersion scales as $\sigma_{\rm H}^2 \sim \langle v^2 \rangle \sim G M_{\rm BH} / r$ --- the relaxation time is
\begin{align}
    T_{\rm 2b} \sim \frac{\sigma_{\rm H}^3}{\mathrm{G}^2 m_{\rm H} \rho_{\rm H}} \propto r^{\alpha - 3/2}\,.
\end{align}
Over a single relaxation time, the entire binding energy of the stellar system is efficiently redistributed. The total binding energy enclosed within radius $r$ scales as $N(r) E(r) \propto r^{-\alpha + 2}$, where $N(r)$ is the number of stars with radius less than $r$ and $E(r)$ is the binding energy of a single star at $r$, implying that the energy flux through radius $r$ is \citep[][\S7.5.9]{Binney_2008}
\begin{align}
    F_{E,{\rm H}} \sim \frac{N(r)E(r)}{T_{\rm 2b}} \propto r^{7/2 - 2\alpha}\,.
\end{align}
In steady state, the distribution function of heavy stars, $f_{\rm H}$, obeys $\dd f_{\rm H}/\dd t = 0$, and thus the energy flux must be radius-independent. This condition yields the classic BW solution $\alpha = 7/4$ \citep{Peebles_1972, Bahcall_1976, Bahcall_1977,Lightman_Shapiro1978}.
In contrast, during strong mass segregation, the heavy stars are sufficiently rare that they  primarily interact with the background of lighter stars rather than with each other. In this regime, dynamical friction dominates over diffusion. Consider a light stellar component with density profile $\rho_{\rm L} \propto r^{-\alpha_{\rm L}}$ and a heavy component with $\rho_{\rm H} \propto r^{-\alpha_{\rm H}}$. The light stars behave effectively as a single-mass population, yielding $\alpha_{\rm L} = 7/4$. The acceleration due to dynamical friction acting on the heavy stars due to the light background obeys \citep{Chandrasekhar_1943}
\begin{align}
   a_{\rm DF} \propto \frac{G^2 m_{\rm H} \rho_{\rm L}}{v^2}\,. 
\end{align}
The corresponding torque on the heavy stars is $\dot{L}_{\rm H} = m_{\rm H}a_{\rm DF} r_{\rm H}$, and since $L_{\rm H} = r_{\rm H} v_{\rm H}$ with $v_{\rm H} \propto r_{\rm H}^{-1/2}$, this implies
\begin{align}
   \dot{r}_{\rm H} \propto -r_{\rm H}^{5/2 - \alpha_{\rm L}}\,. 
\end{align}
Requiring the inward number flux $4\pi r_{\rm H}^2n_{\rm H}|\dot r_{\rm H}|$ to be independent of radius in steady state then gives $\alpha_{\rm H}=9/2-\alpha_L=11/4$ \citep{Alexander_2009, Keshet_2009, Linial_Sari2022}.
The $r^{-3/2}$ DM steady state exhibits much lower densities than the $r^{-7/3}$ slopes predicted by \citet{Gondolo_1999}. Consequently, the time-scales on which these relaxed profiles are realised will indicate how long these initially steep spikes can be maintained for in the stellar sphere.

\subsection{Time-scales of our System}
\label{subsec:Time-scales}
The time to reach these steady-state profiles is some multiple of the relaxation time, which can be quite long. We now examine this time-scale in the presence of a multi-mass stellar population. Several dynamical processes, each characterised by its own time-scale, are at play within the SoI of an MBH \citep[e.g.,][]{Kocsis_2011}. Of primary interest are the two non-resonant relaxation time-scales: two-body relaxation, $T_{\rm 2b}$, and dynamical friction, $T_{\rm DF}$ --- both naturally captured within the FP framework. In a multi-mass stellar cusp, these must be computed by averaging over all particle species.

The two-body relaxation time for a star of species $i$ with velocity $v$ may be defined as
\begin{align}
  T_{{\rm 2b},i} \sim \frac{v^2}{\dd \langle (\Delta v)^2 \rangle/\dd t}\,,  
\end{align}
where $\langle (\Delta v)^2 \rangle$ is the cumulative mean-squared change in velocity due to weak encounters with stars of all species. 
Consider an encounter between this star and a star of species $j$, with mass $m_j$, relative velocity $v_{\rm rel}$, and impact parameter $b$. The resulting velocity kick scales as $\Delta v \sim 2 \mathrm{G} m_j/(bv_{\rm rel})$ \citep[][Eq.~(8.41)]{Binney_2008}. Taking the time average and integrating over all impact parameters and over a mass distribution with number density $n(m)$ gives
\begin{align}
  \frac{\dd \langle (\Delta v)^2 \rangle}{\dd t} \propto \frac{G^2 \ln \Lambda}{v} \int n(\tilde{m}) \tilde{m}^2 \dd \tilde{m}\,.
\end{align}
For a discretised multi-mass spectrum with stellar masses $m_j$, the number density per unit mass reads
\begin{align}
    n(m) = \sum_j \frac{\rho_j}{m_j} \delta(m - m_j)\,,
\end{align}
where $\rho_j$ is the mass density of species $j$. Substituting this into the standard expression for the relaxation time of a single stellar component \citep[][Eq.~(7.106)]{Binney_2008} yields the two-body relaxation time for species $i$ in a discretised multi-mass spectrum (see \footnote{Different definitions of the relaxation time lead to slightly different numerical pre-factors; see, e.g., \citet{Gnedin_2004, Merritt_2004}.}):
\begin{align}
    T_{{\rm 2b}, i} = \frac{0.34 \sigma_i^3}{\mathrm{G}^2 \ln \Lambda \int n(m) m^2 \dd m} = \frac{0.34 \sigma_i^3}{\mathrm{G}^2 \ln \Lambda \sum_{j} \rho_j m_j}\,, 
    \label{eq:t_2b rel}
\end{align}
where $\sigma_i$ is the velocity dispersion and $\ln \Lambda$ is the Coulomb logarithm.
A similar argument applies to dynamical friction. The acceleration $a_{\rm DF}$ experienced by a star of species $i$ due to dynamical friction from species $j$ scales as $\rho_j$, and summing over all species gives the net dynamical friction acceleration. Using $T_{\rm DF} = v/a_{\rm DF}$, we obtain \citep[Eq.~(5.32)]{Merritt_2013_book}
\begin{align}
    T_{{\rm DF}, i} &= \frac{0.945}{Q} \frac{\sigma_i^3}{\mathrm{G}^2 m_i \ln \Lambda \sum_{j} \rho_j} \label{eq:t_DF rel}\,,\\
    Q &= \frac{12}{2\sqrt{2}}\left[\int_0^{1/\sqrt{2}} e^{-t^2}\dd t - \frac{1}{\sqrt{2}}e^{-1/2}\right] \approx 0.747\,,
\end{align}
where, in evaluating $Q$, we have assumed all particles move at a velocity equal to that species' velocity dispersion \footnote{A more thorough analysis yields only order unity corrections \citep{Merritt_2013_book}.}. From Eqs.~\eqref{eq:t_2b rel} and~\eqref{eq:t_DF rel}, we find 
\begin{align}
    \frac{T_{{\rm 2b},i}}{T_{{\rm DF},i}} \approx 0.2688 \frac{m_i}{\langle m \rangle}\,, 
    \label{eq:t_2b/t_DF}
\end{align}
where
\begin{align}
    \langle m \rangle = \frac{\sum_j \rho_j m_j}{\sum_j \rho_j} \label{eq:eff_mass} \,.
\end{align}
The ratio of the relaxation time-scales thus scales with stellar mass. Consequently, in a mass-segregated cusp away from equilibrium, heavier species experience more rapid inward migration via dynamical friction, while lighter species are primarily heated by two-body diffusion. This mass dependence --- combined with the fact that mass segregation pushes heavy species inwards --- suggests that DM particles near the centre relax more rapidly in a multi-mass cusp than in a single-mass one.

\subsection{Stellar Model and DM Initial Density}
\label{subsec:Stellar Model and DM initial density}

We now specify the adopted initial ($t = 0$ in our FP model) conditions for the DM and stellar components, including their radial density profiles, density normalisations, and the stellar mass function.

\subsubsection{Initial DM Density Profiles}

For the DM component, we assume an initial power-law profile
\begin{align}
  \rho_{\rm DM}(r) = \rho_{\rm sp} \left(\frac{r}{r_{\rm sp}}\right)^{-\gamma_i},
\end{align}
where $\rho_{\rm sp}$ is the density at $r = r_{\rm sp}$ and we have set the initial slope to $\gamma_i = 7/3$, as predicted by \citet{Gondolo_1999} for an MBH formed inside a pre-existing NFW DM cusp. To obtain the normalisation, the standard convention in the literature is to either (i) estimate the DM density at the radius of the Sun and extrapolate inwards to the spike radius $r_{\rm sp}$ assuming an NFW profile \citep{Gondolo_1999}, or (ii) model a high-redshift NFW halo with virial mass $10^6 \, \Msolar$, use known concentration–mass relations, and match to the NFW density \citep{Eda_2015}. We take the matching radius to be $r_{\rm sp} = 0.2 r_{\rm SoI}$ \citep{Gondolo_1999}. Both approaches yield nearly identical values of $\rho_{\rm sp} = 226 \, \Msolar/\pc^3$.

\subsubsection{Initial Stellar Density Profiles}
\label{subsec:Initial Stellar Density Profiles}

We assume the stars are initially distributed according to a spherically symmetric Hernquist profile, typically used to model galactic bulges, given by
\begin{align}
    \rho_*(r) = \rho_{0,*} \left(\frac{r}{r_0}\right)^{-1} \left[1 + \frac{r}{r_0} \right]^{-3}\,,
    \label{eq:Hern stellar profile}
\end{align}
where $\rho_{0,*}$ is a normalisation constant, and $r_0$ is given by \citep{Hernquist_1990, Kormendy_2013, Hon_2022}
\begin{align}
    r_0 = 17.853 \left(\frac{\mSMBH}{10^6 \, \Msolar}\right)^{0.748} \pc\,.
\end{align}
Although the Hernquist profile scales as $r^{-1}$ inside the SoI \footnote{This is because $r_{\rm SoI} \ll r_0$ with approximate equality only when $\mSMBH \approx 5\, \Msolar$. This follows from combining the $R_e$--$M_{\rm bulge}$ relation \citep{Hon_2022}, the $M_{\rm bulge}$--$\mSMBH$ relation \citep[][Eq.~(10)]{Kormendy_2013} and $r_{\rm SoI} = G\mSMBH/\sigma^2$ \citep[][Eq.~(7)]{Kormendy_2013}.} --- the region of primary interest --- we retain the full profile to ensure that the inner region evolves self-consistently with the surrounding bulge, which acts as a reservoir of stars and sets the outer boundary condition of the SoI. 
While the Hernquist profile is widely used, different NSC formation scenarios, such as via in situ star formation \citep{Bartko_2009, Bartko_2010} or the inspiral of globular clusters \citep{Antonini_2012, Antonini_2013}, predict different initial, pre-relaxation, density profiles. For completeness, we discuss the impact of these different profiles on our results in Appendix \ref{appsec:Diff NSC ICs}.
We normalise the stellar density profile such that the total stellar mass within the SoI is equal to the MBH mass,
\begin{align}
    \mSMBH =
    \int_0^{r_{\rm SoI}}4\pi r^2 \rho_*(r)\; \dd r \label{eq:M_BH norm}
\end{align}
where
\begin{align}
    r_{\rm SoI} = 1.468 \left(\frac{\mSMBH}{10^6 \, \Msolar}\right)^{0.544} \pc\,, \label{eq:RoI eqn}
\end{align}
which comes from the definition $r_{\rm SoI} = \mathrm{G}\mSMBH/\sigma^2$ and the famous $M$--$\sigma$ relation given in \citet[Eq.~(7)]{Kormendy_2013}. 
This total mass density is then distributed among individual stellar components according to the stellar mass function given next.

\subsubsection{Stellar Mass Function}
\label{subsec:SMF}
We adopt the stellar mass function of \citet[][Eq.~6]{Kroupa_2001}, parametrised as a broken power-law,
\begin{align}
    \mathcal{N} \propto
    \begin{cases}
        \left(m/\Msolar\right)^{-0.3}\,, & \text{for } m \leq 0.08\,\Msolar\,; \\
        \frac{2^{3/2}}{125}\left(m/\Msolar\right)^{-1.8}\,, & \text{for } 0.08\,\Msolar \leq m < 0.5\,\Msolar\,; \\ 
        \frac{2^{3/5}}{125} \left(m/\Msolar\right)^{-2.7}\,,& \text{for } 0.5\,\Msolar \leq m < 1\,\Msolar\,; \\
        \frac{2^{3/5}}{125} \left(m/\Msolar\right)^{-2.3}\,,& \text{for } 1\,\Msolar \leq m \,.
    \end{cases}
\end{align}
Stellar-mass BHs also contribute to the total mass budget. However, they are expected to comprise only $\mathcal{O}({3})\%$ of the total stellar mass within the Milky Way's SoI \citep{OLeary_2009, Alexander_2009, Amaro_Seoane_2011}. Moreover, their number density, mass distribution, and radial distribution are poorly constrained. We therefore neglect their contribution to the evolution of the stellar sphere \footnote{The sBH mass function is generally expected to be top-heavy; including it would shorten relaxation times and thus only strengthen our conclusions.}.

We normalise $\mathcal{N}(m)$ so that the total stellar mass within the SoI satisfies
\begin{align}
    \mSMBH = \int_{m_{\rm min}}^{m_{\rm max}} \mathcal{N}(m) m \dd m\,. \label{eq:MF total mass}
\end{align}
For the numerical implementation, we discretise the mass function into $N_s$ logarithmically spaced mass components $m_i$, with $i \in \{1,2,\ldots, N_s\}$. The continuous distribution is approximated as
\begin{align}
    \mathcal{N}(m) = \mathcal{N}_0 \sum_i \delta (m - m_i) \mathcal{N}(m_i) \,,
\end{align}
where $\mathcal{N}_0$ is a normalisation constant. The mass assigned to the $i^{\rm th}$ stellar component is then
\begin{align}
    M_i = \mSMBH \, \dfrac{\int_{m_{i,{\rm lower}}}^{m_{i,{\rm upper}}} m \, \mathcal{N}(m) \, \dd m}{\sum_j \int_{m_{j,{\rm lower}}}^{m_{j,{\rm upper}}} m \, \mathcal{N}(m) \, \dd m}\,,\label{eq:stellar comp masses}
\end{align}
where $m_{i,{\rm lower}}$ and $m_{i,{\rm upper}}$ denote the lower and upper edges of each mass bin, taken to be the logarithmic midpoints between adjacent discrete masses. 
Our goal is to compare a single-effective-mass model with a multi-mass model. Since $\mathcal{N}(m) \propto m^{-2.3}$ for $m > 1\,\Msolar$, the effective mass (Eq.~\ref{eq:eff_mass}) is top-heavy and therefore very sensitive to the upper mass cut-off. However, the stellar mass function for stars heavier than $10 \, \Msolar$ is poorly constrained, and such stars have short lifetimes $\lesssim 30$ Myr \citep[][Tab. A.1]{Eggenberger_2021}. Thus, we consider two multi-mass cases:
\begin{itemize}
    \item A \textit{truncated} mass function, with masses ranging from $0.1\,\Msolar$ to $10\,\Msolar$.
    \item An \textit{extended} mass function, with masses ranging from $0.1\,\Msolar$ to $100\,\Msolar$.
\end{itemize}
Each model contains ten stellar species logarithmically spaced between the lower and upper mass limits. In the multi-mass models, all species share the same initial radial profile (Eq.~\ref{eq:Hern stellar profile}), but with different normalisations determined by Eq.~\eqref{eq:stellar comp masses}. In the corresponding single-effective-mass models, which have masses $1.6\,\Msolar$ (truncated) and $7.55\,\Msolar$ (extended), the density profiles are also normalised such that the total mass inside the SoI is $\mSMBH$. We have computed the effective masses from the initial stellar mass function and density normalisation described above.

\subsubsection{Boundary Conditions and the Loss Cone}
\label{subsubsec:BCs and LC}
\begin{figure*}
    \centering
        \subcaptionbox{Multi-mass stellar model with ten species logarithmically spaced between $0.1\,\Msolar$ and $100\,\Msolar$.
        \label{subfig:DM_sim10_threepanel_heavy}}
        {\includegraphics[width=0.99\linewidth]{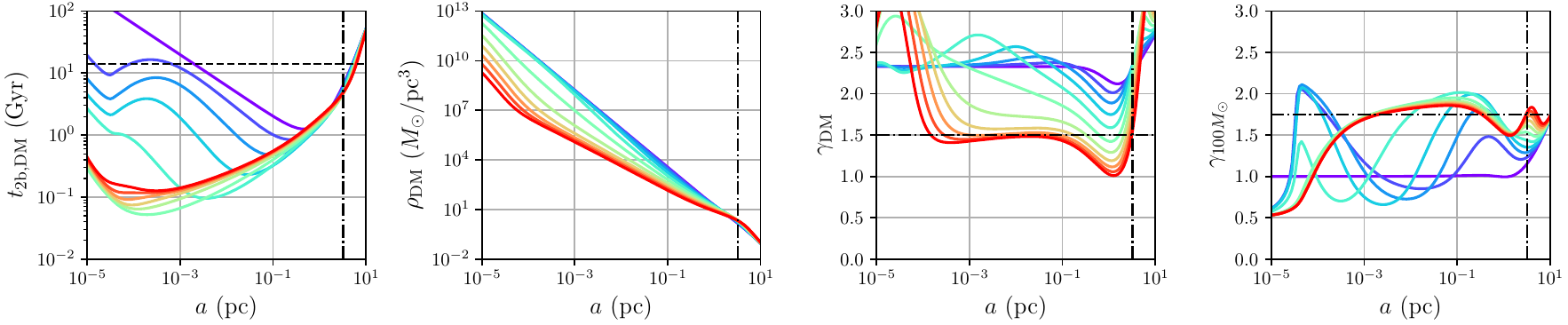}}\\\vspace{0.25cm}
        \subcaptionbox{Single-mass stellar model with mass $7.55\,\Msolar$.
        \label{subfig:DM_sim1_threepanel_heavy}}
        {\includegraphics[width=0.99\linewidth]{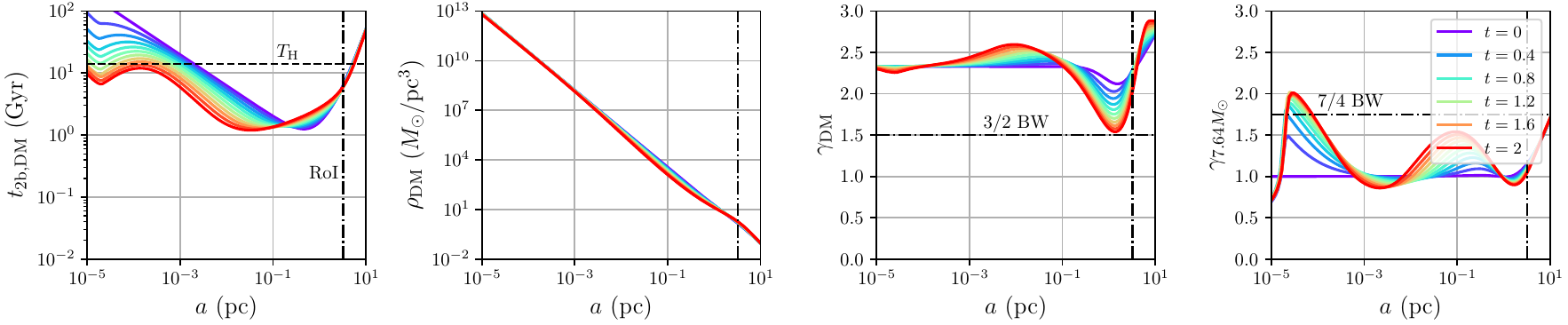}}
    \caption{
    Time evolution of the DM and stars, whose mass function follows our extended model, around an MBH of mass $4.3 \times 10^6 \, \Msolar$, integrated over 2 Gyr. From \textit{left} to \textit{right}, the panels show:~the DM two-body relaxation time-scale, the DM density profile as a function of radius, the DM density power-law slope $\gamma_{\rm DM} = -\mathrm{d}\ln \rho_{\rm DM}/\mathrm{d}\ln r$, and the density power-law slope of the most massive stellar species. The different colours represent evolution time in units of Gyr as indicated in the legend in the bottom right panel. The \textit{top} and \textit{bottom} rows display our multi-mass model and single-effective-mass model, respectively. The vertical black dot–dashed line marks the MBH radius of influence, the horizontal dashed line in the left panels denotes one Hubble time (14 Gyr), and the horizontal dot–dashed lines in the right-hand panels denote the relaxed BW profiles. Clearly, the multi-mass model exhibits significantly faster relaxation compared to the single-mass case.
    }
    \label{fig:DM_threepanel_plots_heavy}
\end{figure*}
We now turn to discussing the boundary conditions of the FP equation. 
We impose an inner cut-off radius in the FP model for each species (stars and DM) inside which objects are assumed to be swallowed by the MBH and removed from the system. \phaseflow approximates the effect of this loss cone, which is in two-dimensional energy-angular momentum space, through a 1D energy-dependent loss term, responsible for the steady-state flux in the angular momentum direction (see \citet[][\S{3}]{Vasiliev_2017}).
For stars, the natural choice is the tidal disruption radius \citep{Hills_1975}
\begin{align}
    r_{\rm tid} = R_{\rm star} \left(\frac{\mSMBH}{m}\right)^{1/3}\,,
\end{align}
where $R_{\rm star}$ and $m$ are the stellar radius and mass, respectively. We will use the stellar mass-radius relation found by \citet{Demircan_1991},
\begin{align}
    \frac{R_{\rm star} (m)}{\Rsolar} &=
    \begin{cases}
        1.06 \left(\frac{m}{\Msolar}\right)^{0.945}, &\text{if } m \leq 1.66 \, \Msolar \,,\\
        1.33 \left(\frac{m}{\Msolar}\right)^{0.555}, &\text{if } m > 1.66 \, \Msolar \,,
    \end{cases} \label{eq:mass-radius rel}
\end{align}
in agreement with subsequent studies \citep{Hurley_2000,Torres_2010,Eker_2018}. We are implicitly ignoring asymptotic-giant-branch and red-giant stars, and instead focus on main-sequence stars, which is the dominant component by numbers. Our results are insensitive to this choice within realistic ranges for $R_{\rm star}$.
Since DM particles are not tidally disrupted, we instead adopt the Newtonian capture radius of the MBH as their inner boundary, taken to be $8\mathrm{G}\mSMBH/\mathrm{c}^2$. While full GR predicts a capture radius of $4\mathrm{G}\mSMBH/\mathrm{c}^2$ for parabolic orbits \citep{Cutler_1994}, our FP model is Newtonian, and we therefore adopt $8\mathrm{G}\mSMBH/\mathrm{c}^2$ for consistency. We note that the inclusion of sBHs would affect loss cone dynamics, which is addressed in detail in \S\ref{sec:DM and EMRIs}.

We truncate our FP model at 1 kpc, where we impose a zero-particle flux Neumann boundary condition. In this way, the boundary of the radius of influence is self-consistently evolved with the rest of the bulge.

\subsection{Results --- Dark Matter Densities in Multi-mass Systems}
\label{subsec:Results -- FP}
We are now finally in a position to present the results of the FP simulations, focussing on the impact of mass segregation on the co-evolution of the stellar and DM cusps around an MBH. In particular, we compare a multi-mass stellar model with a single-mass model in which the stellar mass is set equal to the effective mass given in Eq.~\eqref{eq:eff_mass}.
Figure~\ref{fig:DM_threepanel_plots_heavy} shows the evolution of the extended-mass model over 2 Gyr; the top and bottom rows pertain to the multi-mass stellar model and the single-effective-mass model, respectively. From left to right, the panels display: the evolution of the DM two-body relaxation time-scale (Eq.~\ref{eq:t_2b rel}), the DM density profile, the local DM density power-law slope $\gamma_{\rm DM} = -\mathrm{d}\ln \rho_{\rm DM}/\mathrm{d}\ln r$ (initially $\rho_{\rm DM} \propto r^{-7/3}$), and the density slope of the most massive stellar species (initially Hernquist; Eq.~\ref{eq:Hern stellar profile}).
In the multi-mass case, the two-body relaxation time is shorter than the single-mass relaxation time by a factor of $\mathcal{O}$(10--100), in agreement with previous work \citep[e.g.,][]{Perets+2007,OLeary_2009}. This is driven by mass segregation: massive stars migrate inwards while lighter stars move outwards, increasing the central density and thereby enhancing two-body scattering. At $t = 0$, the relaxation times are identical in both models by construction, since they share the same initial sum-squared mass and the multi-mass model has not yet had time to segregate.
These differences affect the DM evolution strongly. In the single-mass case, the initial $\rho_{\rm DM} \propto r^{-7/3}$ profile is relatively well-maintained over the 2 Gyr integration time. In contrast, in the multi-mass model the DM cusp relaxes to the BW profile $\rho_{\rm DM} \propto r^{-3/2}$ (indicated by the horizontal dash–dotted line in the centre-right panel) within one Gyr, down to distances of $a \sim 10^{-4} \pc$. At smaller distances, stellar densities are reduced due to tidal disruption, leading to an increase in relaxation times, as seen in the far left panel of Figure~\ref{subfig:DM_sim10_threepanel_heavy}, and hence an inability for the spike to relax completely \footnote{But of course, sBHs and other compact objects (which we have ignored in the sense that the loss-cone boundary is the stellar disruption radius, for all $m$) should still be there, and could impact relaxation rates --- however this would be mild due to their low number densities.}. Thus, despite DM densities continuing to decrease at these small radii (as seen in the centre-left panel of Figure~\ref{subfig:DM_sim10_threepanel_heavy}), the slope must exceed the initial $\gamma_i = 7/3$ for $\rho_{\rm DM}$ to remain continuous.
Similarly, the heaviest ($100 \, \Msolar$) stellar component relaxes towards the weak $\rho_* \propto r^{-7/4}$ BW solution only in the multi-mass model (far right panels of Figure \ref{fig:DM_threepanel_plots_heavy}). Here the number densities of heavy stars are sufficiently large to put the system in the weak mass segregation regime, rather than the strong one.
The DM density is reduced by a factor of $10^{3}$--$10^{5}$ after $t \sim$ 0.8--1.6 Gyr in the central region. Even after first reaching the $3/2$ BW profile, the DM density continues to decline due to ongoing heating by stellar encounters. This demonstrates that multi-mass stellar models cause DM densities to drop by orders of magnitude far earlier than in single-effective-mass models, and are thus essential for accurately modelling the long-term evolution of DM spikes. 
We show the evolution of the truncated-mass-function model in Appendix \ref{appsec:Evolution with the Truncated Mass Function}. The qualitative behaviour is the same as in the extended case but proceeds on longer time-scales. Further, as stated in \S\ref{subsec:Initial Stellar Density Profiles}, different NSC formation histories lead to different initial density profiles. In Appendix \ref{appsec:Diff NSC ICs}, we discuss how the difference in relaxation times between multi-mass and single-mass models varies for these different profiles. We indeed find that a reduction in relaxation times for multi-mass models is a generic outcome.
Thus, realistic stellar populations drive DM spikes toward a lower-density profile in the stellar sphere on $\mathcal{O}(\Gyr)$ time-scales. However, we emphasise that the innermost regions are not well described by our FP model and require separate treatment, which we turn to next.
\section{Evaporation of DM Spikes by EMRIs}
\label{sec:DM and EMRIs}
As discussed in \S\ref{subsec:FP and SSE}, the FP formalism is no longer valid inside the loss cone, where sBHs play an important dynamical role and must therefore be reintroduced into our model. To do so, we assume that any sBHs of mass $\mEMRI$ follow the same density profile as the stellar component of the same mass.
Inside the loss cone, stellar and sBH number densities are very low, while the DM number density remains high due to the extremely small individual DM particle mass. There, the evolution of the DM distribution is governed by sporadic, discrete interactions with individual sBHs. The large number of DM particles ensures that strong DM--sBH encounters occur at an appreciable rate --- much more frequently than strong star--star, star--sBH, or sBH--sBH encounters.
It is well-known that individual objects generically heat DM particles, transferring them to larger radii rather than driving an inward flux \citep{Kavanagh_2020, Coogan_2022, Mukherjee_2024, Kavanagh_2024, Alonso_Alvarez_2024, Karydas_2024, Tiruvaskar_2025}. While DM particles can in principle undergo two-body relaxation with one another, the associated time-scale (Eq.~\ref{eq:t_2b rel}) is extremely long --- $\mathcal{O}(10^{70})$ yr \citep{Kavanagh_2020} --- because of the small DM particle mass. Consequently, once a DM particle is ejected to low binding energy or removed from the system, it cannot be replenished by inward relaxation, making any EMRI-driven depletion irreversible.
LISA is expected to detect EMRIs predominantly at redshifts $z \lesssim 2$--$3$ \citep{Babak_2017, Chapman_2025, Speri_2026}. An EMRI occurring at $z = 3$ takes place when the Universe was 2.14 Gyr old (for \emph{Planck} \citep{Planck:2018vyg} cosmological parameters). This sets the minimum time available for DM spike evolution among detectable events. We therefore adopt a conservative approach and model the cumulative spike evolution over the first $2.14$ Gyr.
\begin{figure}
    \centering
    \includegraphics[width=0.49\textwidth]{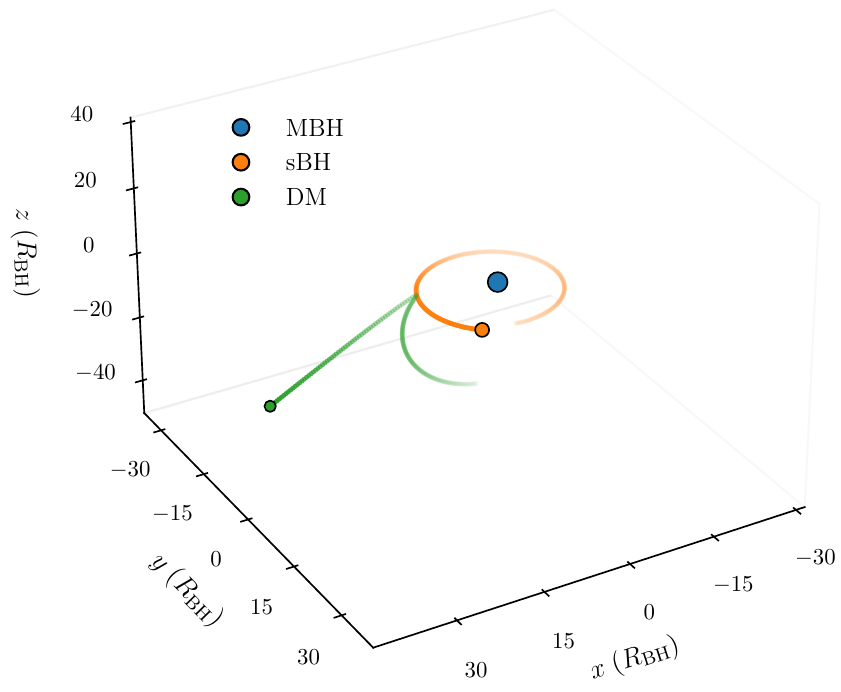}
     \caption{Example three-body evolution showing a  DM particle being ejected via a gravitational slingshot interaction with an EMRI.}
    \label{fig:triple_evolution_last_frame_fixed}
\end{figure}
Here, we develop a three-body scattering model and evolve it numerically through $N$-body simulations with post-Newtonian (PN) corrections. We use this to quantify the gradual evaporation of the inner DM spike, caused by repeated strong encounters between DM particles and EMRIs.
\subsection{Our Three-Body Model}
\label{subsec:Our 3b Model}
Since collisionless DM particles behave independently of each other (due to their total mass being much smaller than the MBH mass), we can model the spike's response to an EMRI by studying the response of a \textit{single} DM particle statistically. Concretely, we repeatedly simulate three-body MBH--sBH--DM encounters over one EMRI orbit.
We deliberately follow a first-principles approach. We focus on the fraction of DM particles that are either ejected from the system or swallowed by the MBH or sBH, rather than evolving the full DM distribution function, since removal from the inner regions is rapid and --- as discussed --- irreversible. If a DM particle is ejected with velocity $v \geq v_{\rm esc} \approx \sqrt{2\mathrm{G}\mSMBH/r}$ (Newtonian approximation), it will have left the SoI within $\mathcal{O}(100)\,\mathrm{kyr}$ --- far shorter than any dynamical relaxation time-scales of our system.
We assume a mass ratio $q = \mEMRI/\mSMBH \ll 1$, so that the sBH acts as a test particle in the MBH potential, and $\mDM \ll \mEMRI$, so that the DM particle is a test particle with respect to both BHs.
We consider DM depletion at very small radii and therefore model the DM particle's orbit around the MBH using post-Newtonian dynamics rather than Newtonian gravity.  We evolve the three-body encounters using \multistar\citep{multistar_ref}, an $N$-body code written in {\tt Fortran}, which includes pairwise post-Newtonian dynamics of order 3.5 (3.5PN) corrections and is hence well suited to this regime. Incorporating relativistic dynamics requires care in defining both the initial orbital parameters and the criteria for particle ejection.  We discuss these details in Appendix~\ref{appsec:Details of Three-Body Simulations}.  Figure~\ref{fig:triple_evolution_last_frame_fixed} illustrates a representative interaction in which a DM particle is ejected through a gravitational slingshot with the EMRI.

Due to the spherical symmetry of the Schwarzschild metric, only the \textit{relative} orientation of the two orbits matters. We therefore obtain a statistical understanding of a DM particle's ejection, for a given periapsis and eccentricity, by fixing the sBH's initial phase and orbital orientation while randomising the DM particle's orientation by sampling its inclination uniformly in $\cos \iota$, as well as uniformly randomising its argument of periapsis and longitude of the ascending node. Modelling the cumulative effect over many repeated EMRIs requires an understanding of EMRI parameter distributions, which we obtain from our FP study in \S\ref{sec:DM Spikes in Mass Segregated Stellar Cusps}, as we describe next.

\subsection{MBH and EMRI Parameter Distributions}
\label{subsec:Param Distributions}
%
We now describe the distributions adopted for the MBH and EMRI parameters. 
\textbf{MBH Mass Range:} MBH masses span a vast range of $\mSMBH \in \left[10^{3},10^{11}\right] \, \Msolar$. Those with $\mSMBH \gtrsim 10^6$--$10^7\,\Msolar$ are likely to have experienced at least one major merger in their past \citep{Conselice_2006, Fakhouri_2010, Ravi_2015, Sijacki_2015, Rodriguez_Gomez_2016, Weinberger_2017, OLeary_2021}. Because major mergers disrupt, or even erase, DM spikes efficiently \citep{Ullio_2001, Milosavljevi_2001, Merritt_2002, Kavanagh_2018}, we set $10^7 \, \Msolar$ as our upper mass limit. Further, nuclear clusters do not necessarily exist around more massive BHs \citep[e.g.,][and references therein]{Neumayer_2020,Hoyer_2021}.
We also impose a lower MBH mass cut-off of $10^4\, \Msolar$ as the understanding of lighter IMBHs, with $\mSMBH \lesssim 10^4\, \Msolar$, remains highly uncertain. Such objects have not yet been observed, so the nature of their host environments is unknown, and their formation channels --- whether via direct collapse, adiabatic growth, or hierarchical mergers --- are still debated, as is the influence they have on surrounding matter \citep{Portegies_Zwart_2004, Mezcua_2017,Guolo_2026}. Coincidentally, this leaves us with almost the exact mass range of the MBHs that are expected to host the EMRIs that LISA will detect \citep{Babak_2017}.
\textbf{EMRI Mass Distribution:} The mass spectrum of IMRIs/EMRIs is poorly constrained, but is generally believed to be dominated by sBHs of mass 10--30$\,\Msolar$, or even heavier due to mass segregation \citep{Barack_2004, Hopman_2009, Aharon_2016, Babak_2017}. We conservatively assume that the largest EMRI mass is $10\,\Msolar$.
\textbf{EMRI Rates:} Despite extensive efforts, predictions for EMRI rates span a broad range. Most studies estimate rates between 1--1000$\,\Gyr^{-1}$ per galaxy, dominated by sBHs \citep{Hils_1995, Sigurdsson_1997, Ivanov_2002, Hopman_2005, Amaro_Seoane_2011, Aharon_2016, Babak_2017, Amaro_Seoane_2018, Panamarev_2019, V_zquez_Aceves_2021, Naoz_2022, Broggi_2022, Rom_2024}. Given these large uncertainties, we treat the EMRI rate as a free parameter and model only the EMRI eccentricity and periapsis distributions. We note, however, that a majority of these works predict sBH EMRI rates of $\mathcal{O}(100-300)\, \Gyr^{-1}$ per MBH.
\textbf{EMRI Eccentricity and Periapsis Distribution:} 
We use \phaseflow to extract a distribution of EMRI eccentricities and periapsides. In particular, since we are assuming all EMRIs to have mass $\mEMRI=10\, \Msolar$, with no heavier contributions, we take the EMRI distribution to be that of the $10\, \Msolar$ stellar component at $t = 2.14\,\Gyr$ of our multi-mass truncated-mass-function FP model run (see Appendix \ref{appsec:Evolution with the Truncated Mass Function}). That is, we approximate the EMRI distribution at earlier times by that at $t = 2.14\,\Gyr$. This is strictly only valid after the system has relaxed, as the EMRI distribution is effectively unchanging --- before this, the EMRI distribution would be skewed to lower periapsides and hence depletion would be reduced. However, incorporating this effect would significantly increase the computational cost of our calculations and only modify the required EMRI rates by at most $\lesssim 30\%$ across nearly the entire MBH mass range considered.

As discussed in \S\ref{sec:Introduction}, our FP model does not describe the region inside the loss cone accurately, where strong interactions dominate the dynamics. However, it does capture the flux into the loss cone accurately. For sBHs, the loss cone is defined by the condition that the GW inspiral time-scale $T_{\rm GW}$ is shorter than the angular-momentum diffusion time-scale, $T_J$. At fixed periapsis $\rp$, this defines the maximum eccentricity, $e_{\rm LC} (\rp)$, an sBH can have before entering the loss cone, which is implicitly determined by $T_{\rm GW}(\rp,e_{\rm LC}) = T_J(\rp,e_{\rm LC})$. Although the loss cone is a two-dimensional region in energy-angular-momentum space, our FP treatment only evolves the energy distribution, with the eccentricity distribution fixed to a thermal one. When angular-momentum diffusion occurs on a time-scale much shorter than energy diffusion, the effect of the loss cone can be incorporated into a one-dimensional model through an energy-dependent sink term, $\nu(E)$. The corresponding loss rate may be written as \citep{Merritt_2013,Vasiliev_2017}
\begin{align}
    \nu(E) = \frac{T_J^{-1}}{\left(u^2 + u^4\right)^{1/4} + \ln(1/\mathcal{R}_{\rm LC})}\,, \label{eq:nu(E)}
\end{align}
where $\mathcal{R}_{\rm LC} = 1-e_{\rm LC}^2$, and
\begin{align}
    u = \frac{1}{\mathcal{R}_{\rm LC}}\frac{P(E)}{T_J}\,.
\end{align}
This is the loss-cone \emph{filling factor}, representing the ratio of the mean-square change of angular momentum per orbital period to the width of the loss-cone boundary, where $P(E)$ is the orbital period.
Due to the dynamics outside the loss cone being governed by weak, stochastic two-body encounters, any sBH that reaches the loss cone boundary will have undergone a random walk to arrive there, as opposed to being kicked into the loss cone by a single strong encounter. Once it crosses the boundary, we assume that its evolution is dominated by GW emission --- by definition of the loss cone. As a result, we take the initial eccentricity for any EMRI with initial periapsis $\rpEMRIz$ to be $\eEMRIz = e_{\rm LC} (\rpEMRIz)$. This allows us to obtain the initial periapsis distribution of EMRIs, $f_{\rpEMRIz}(\rpEMRIz)$, through
\begin{align}
    f_{\rpEMRIz} &= 
    \begin{cases}
        \displaystyle
        \frac{\nu \, f^{(10)}_{\rpEMRIz}}{\int_{T_{\rm GW} = T_J} \nu \, f^{(10)}_{\rp} (\rp) \dd \rp}\,, &\text{if  }\,\,\, T_{\rm GW} = T_J\,,\\
        0\,, &\text{else}\,,
    \end{cases} \label{eq:f_EMRI}
\end{align}
where $f^{(10)}_{\rp} (\rp)$ is the periapsis distribution of the $10 \, \Msolar$ stellar component (matching the assumption $\mEMRI=10\, \Msolar$) in the FP model at $e = e_{\rm LC} (\rp)$, and $\nu = \nu(\rp, e_{\rm LC})$.
The GW inspiral time is given by \citep{Peters1964}
\begin{align}
    T_{\rm GW} = \frac{12 c_0^4}{19 \beta} \int_0^{\eEMRI} \frac{\tilde{e}^{29/19}}{(1 - \tilde{e}^2)^{3/2}} \left(1 + \frac{121}{304} \tilde{e}^2\right)^{1181/2299} \dd \tilde{e} \,, \label{eq:GW insp time}
\end{align}
where
\begin{align}
    \beta &= \frac{64 \mathrm{G}^3}{5\mathrm{c}^5} \mSMBH \mEMRI (\mSMBH + \mEMRI) \,,\\
    c_0 &= \frac{\rpEMRI (1 + \eEMRI)}{\eEMRI^{12/19}} \left[1 + \frac{121}{304} \eEMRI^2\right]^{-870/2299}\,.
\end{align}
Note that numerical integration of Eq.~\eqref{eq:GW insp time} becomes difficult for $1 - \eEMRI \ll 10^{-4}$, and so for such values we use the approximation \citep{Peters1964}
\begin{align}
    T_{\rm GW} = \frac{3}{85} \frac{\mathrm{c}^5}{\mathrm{G}^3} \frac{\rpEMRI^4 (1 + \eEMRI)^{7/2}}{\sqrt{1-\eEMRI} \mSMBH \mEMRI (\mSMBH + \mEMRI)}\,, \label{eq:T_GW}
\end{align}
which holds when $1 - \eEMRI \ll 1$. 
Let us now compute $T_J$, to finalise specifying $f_{\rpEMRIz}(\rpEMRIz)$ in Eq.~\eqref{eq:f_EMRI}. 
The characteristic time to alter an orbit's angular momentum $J$ by order of itself is \citep{Merritt_2013}
\begin{align}
    T_J = \left(\frac{J}{J_{\rm c}}\right)^2 T_{\rm rel} = (1 - \eEMRI^2) T_{\rm rel}\,,
\end{align}
where $T_{\rm rel}$ is the shortest relaxation time-scale of the system (whether in angular momentum or in energy), $J_{\rm c}$ is the angular momentum of a circular orbit with the same orbital energy and we have used the fact that the specific angular momentum of an orbit is $\propto \sqrt{1 - \eEMRI^2 }$. We hence set $T_{\rm rel}$ as the smaller of the two-body relaxation time (Eq.~\ref{eq:t_2b rel}) and the scalar resonant relaxation (SRR) time 
\begin{align}
    T_{\rm rel} = \min \left[T_{\rm 2b}, T_{\rm SRR}\right]\,. \label{Eq:T_rel}
\end{align}
The latter is given by \citep{Hopman_2006a}
\begin{align}
    T_{\mathrm{SRR}} = \frac{A_{\rm RR}(e) \mathrm{G}^{-1}\mSMBH^2 a^3 }{M_{\rm star, \, enc} + \mSMBH} & \frac{1}{\sum_j M_j M_{j,{\rm \, enc}}} \left|\frac{1}{T_{\rm N}} - \frac{1}{T_{\rm GR}}\right| \,, \label{eq:SRR}
\end{align}
where $M_{j,{\rm \, enc}}(a)$ is the total mass of species $j$ with semi-major axis less than $a$, $M_{\rm star, \, enc}(a) = \sum_j M_{j,{\rm \, enc}}(a)$, $A_{\rm RR}(\eEMRI) = 4/(\pi \eEMRI)^2$ \citep{Gurkan_2007}, and \citep{Einstein_1915, Hopman_2006a}
\begin{align}
    T_{\rm GR} &= \frac{\mathrm{c}^2 a^{5/2}(1-\eEMRI^2)}{3 (\mathrm{G}\mSMBH)^{3/2}} \,,\\
    T_{\rm N} &= A_N\frac{\mSMBH}{M_{\rm star, \, enc}(a)} P(a) \,,
\end{align}
are the  prograde relativistic apsidal advance and the retrograde Newtonian apsidal advance time-scales, respectively. For simplicity, we adopt $A_N = 1/\pi$, corresponding to circular orbits. Although $A_N$ increases with eccentricity \citep{Kocsis_Tremaine2015}, our results are insensitive to this because $T_{\rm GR} \ll T_{\rm N}$ for EMRIs and thus increasing $T_{\rm N}$ will leave $T_{\mathrm{SRR}}$ practically unchanged. 

\begin{figure}
    \centering
    \includegraphics[width=0.49\textwidth]{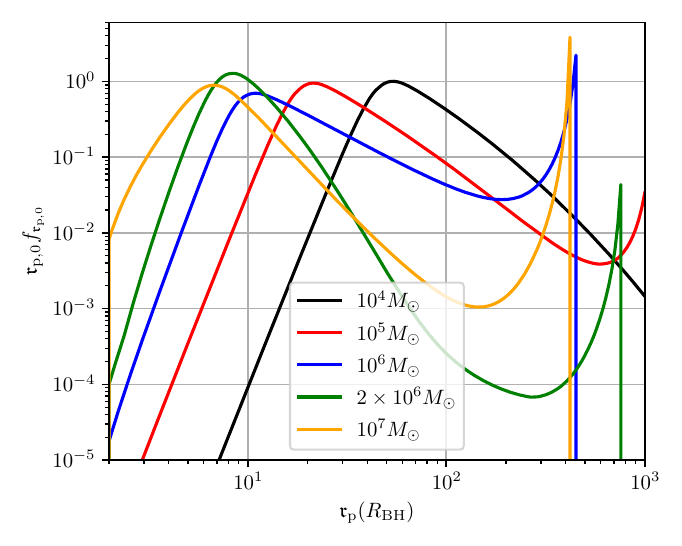}
     \caption{The distribution of initial EMRI periapsis (Eq.~\ref{eq:f_EMRI}) per logarithmic interval for $\mSMBH = [10^4, 10^5, 10^6, 2 \times 10^6, 10^7]\,\Msolar$ at $t = 2.14$ Gyr. The left cut-off is determined by either $\rpEMRIz > 2\RSMBH$ or the maximum semi-major axis, taken to be the RoI of the MBH. The right cut-off arises as there does not exist any $0 \leq e_{\rm LC}< 1$ such that $T_{\rm GW} = T_J$, i.e. the GW-induced inspiral is too slow compared to phase space diffusion.}
    \label{fig:f_rp_dists}
\end{figure}

We can see that faster apsidal precession shortens the coherence time of resonant torques, thereby increasing SRR time-scales, as described in \S\ref{subsec:FP and SSE}. Note that while Eq.~\eqref{eq:SRR} formally implies $T_{\mathrm{SRR}} \to 0$ when $T_{\rm N} = T_{\rm GR}$, this behaviour is an artefact of the framework used, which models orbits as ``fixed wires''. In reality, the relaxation time does not vanish but instead saturates at $T_{\rm SRR} \sim T_{\rm VRR}$, where $T_{\rm VRR}$ is the VRR time-scale.
Due to the strong dependence on $T_J$, the EMRI distribution will vary significantly depending on whether or not the stellar system has relaxed by $t = 2.14$ Gyr. Relaxed cusps will have ``forgotten'' their initial conditions and will follow the relaxed BW profile. However, the state of non-relaxed cusps is strongly dependent on the NSC's initial conditions. We know that $T_{\rm SRR}$ increases with $\mSMBH$ (Eq.~\ref{eq:SRR}) and $T_{\rm 2b}$ scales with $\sigma^3 \propto \mSMBH^{3/2}$ (Eq.~\ref{eq:t_2b rel}). Thus, relaxation times increase with $\mSMBH$ and $\mUR \approx 1.7 \times 10^6 \, \Msolar$ is the critical mass above which they exceed 2.14 Gyr when starting with a Hernquist profile. Due to poor constraints on NSC initial conditions, as discussed in Appendix \ref{appsec:Diff NSC ICs}, we are unable to draw strong conclusions for NSCs whose central MBH mass exceeds $\mUR$. However, for the purposes of having a reference, we still present the results for such systems.
Figure~\ref{fig:f_rp_dists} shows the ``measure-weighted'' initial-periapsis distribution function, or distribution per logarithmic interval given by $\rpEMRIz f_{\rpEMRIz}$ where $f_{\rpEMRIz}$ is given in Eq.~\eqref{eq:f_EMRI}, at $t = 2.14$ Gyr for galactic centres whose central MBH has mass $\mSMBH = 10^4 \, \Msolar, 10^5 \, \Msolar, 10^6 \, \Msolar, 2 \times 10^6 \, \Msolar$, and $10^7 \, \Msolar$. The EMRI distribution changes significantly with the MBH mass, particularly above $\mUR$ where there is a strong skew towards smaller periapsides. This mass dependence is explained in detail in Appendix \ref{appsec:EMRI dis fn}. We have plotted the weighted distribution, $\rpEMRIz f_{\rpEMRIz}$, as opposed to the distribution function $f_{\rpEMRIz}$ alone, to emphasise which regions of parameter space contribute the most EMRIs.
We see clear boundaries on the permissible values of $\rpEMRIz$. The left cut-off corresponds to either $\rpEMRIz > 2\RSMBH$ or the MBH's RoI (the maximum semi-major axis), whereas the right cut-off is set by the condition that $\eEMRI_0 = e_{\rm LC} > 0$ when solving $T_{\rm GW}(\rpEMRIz,\eEMRIz) < T_J(\rpEMRIz,\eEMRIz)$ (see Appendix \ref{appsec:EMRI dis fn} for further details). The eccentricity predictions obtained are in qualitative agreement with previous EMRI studies by \citet{Amaro_Seoane_2007, Sari_Fragione_2019, Broggi_2022, Mancieri_2024, Mancieri_2025}. We apply this procedure consistently across the full range of MBH masses considered.
We have neglected other EMRI formation channels, which we briefly discuss for completeness. Previous work has shown that strong interactions can lead to direct-plunge orbits and may contribute to EMRI rates mildly \citep{Babak_2017, Rom_2024, Zhong_2023,Amaro_Seoane_2025}. Other potential channels include wet EMRI formation in gaseous discs, which are expected to increase EMRI rates significantly for MBHs with an AGN \citep{Levin_2007, Kennedy+2016, Panamarev+2018, Pan_2021, Derdzinski_2023, Sun_2025}, and EMRI production via the Hills mechanism of binary or triple tidal disruption, which, again, is thought to enhance EMRI rates only mildly \citep{Hills_1988, Miller_2005, Sari_Fragione_2019, Raveh_2020}. Although these channels are expected to have different EMRI periapsis and eccentricity distribution functions and therefore modify $f_{\rpEMRIz}$, such channels are sub-dominant and would only serve to strengthen our results.

\subsection{Computing DM Ejection Rates}
\label{subsec:Ejection Rates}

We next derive the DM ejection fractions. We begin by analysing the ejection probability associated with a single EMRI orbit, and its dependence on the mass ratio. We then extend this to compute the cumulative ejection probability over the full lifetime of a single EMRI, and finally over the entire population of EMRIs experienced by an MBH up to $t = 2.14$ Gyr ($z = 3$).

\subsubsection{Scaling with Mass Ratio}
\label{subsubsec:Scaling with Mass Ratio}

%
\begin{figure}
    {\includegraphics[width=0.46\textwidth]{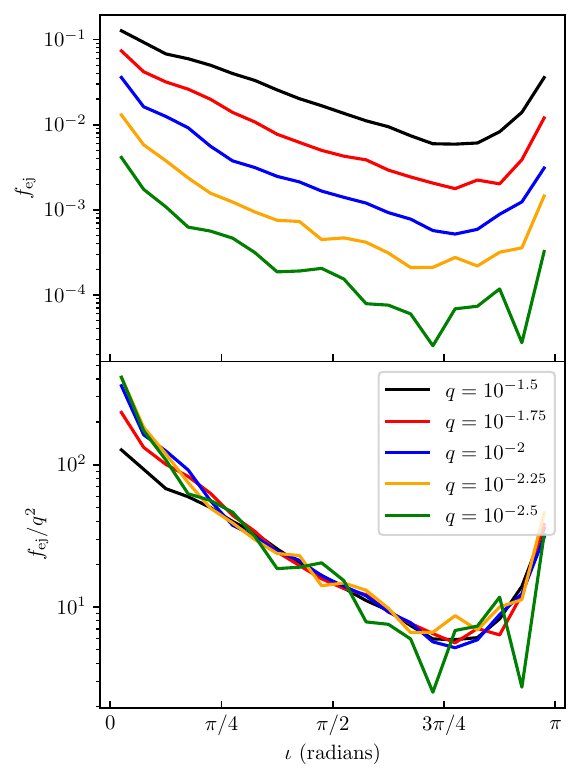}}
     \caption{Ejection fraction of DM particles as a function of inclination for different mass ratios $q$ (\textit{top panel}), and the same curves rescaled by $q^2$ (\textit{bottom panel}). The collapse of the rescaled curves demonstrates the expected $q^2$ scaling (see Appendix~\ref{appsec:Scaling with Mass Ratio}). In all cases we fix $\eDM = 0.2$, $\rpDM = 10\RSMBH$, $\eEMRI = 0.4$, $\rpEMRI = 12\RSMBH$, and $\mSMBH = 10^6 \, \Msolar$ (these ejection fractions are independent of $\mSMBH$ for fixed $q$, as discussed) while randomising the DM particle's orbital orientation and phase.}
    \label{fig:f_ej_scaling}
\end{figure}
We wish to understand how the probability of ejection of a single DM particle over a single EMRI orbit scales with the EMRI's mass ratio, $q$. If this scaling is a simple power-law, then we can use the ejection fractions from a single value of $q \ll 1$ to extrapolate to any value of $q \ll 1$. This would prove very useful in reducing the number of $N$-body simulations we must run.

Simple arguments suggest that the ejection fraction is (explicitly) independent of $\mSMBH$ and instead scales as $q^2$ (see Appendix \ref{appsec:Scaling with Mass Ratio}).  We use \multistar\citep{multistar_ref} to verify this scaling and determine the constant of proportionality.  Now, EMRIs with $10^{-6} \lesssim q \lesssim 10^{-3}$ have ejection fractions of $f_{\rm ej} \sim 10^{-12}$--$10^{-6}$, implying that we would need to run $\mathcal{O}(1/f_{\rm ej})$ simulations to get a reliable estimate for the constant of proportionality, which is unfeasible.  Instead, we perform simulations in the range $10^{-2.5}\leq q \leq 10^{-1.5}$ to confirm the scaling relation.

Figure~\ref{fig:f_ej_scaling} displays the ejection fractions $f_\mathrm{ej}$ and $f_\mathrm{ej}/q^2$ as a function of inclination for several $q$, computed from $2 \times 10^6$ \multistar simulations, per mass ratio, of a single EMRI orbit.  During these runs, the DM particle's orbital parameters are randomised as described in \S\ref{subsec:Our 3b Model}.  We were limited to $q \gtrsim 10^{-2.5}$ due to progressively increasing noise for smaller mass ratios.  From these results, the ratio $f_\mathrm{ej}/q^2$ is clearly independent of $q$, confirming the analytic prediction.  Thus, when calculating ejection fractions per orbit, we take the computed value for $q = 10^{-2}$ and extrapolate down to smaller mass ratios.  While Figure~\ref{fig:f_ej_scaling} shows the ejection fraction specifically for $\mSMBH = 10^6 \, \Msolar$, $\rpEMRI = 12\RSMBH$, $\eEMRI = 0.4$, $\rpDM = 10\RSMBH$, $\eDM = 0.2$,  we find identical $q^{2}$ scaling relations for other parameters.  We explicitly confirmed the $\mSMBH$-independence numerically by comparing ejection fractions for various MBH masses while keeping all other parameters fixed.

\subsubsection{Cumulative Ejection Probability over an MBH Lifetime}

We have established that a given DM {\rm particle in the inner region is ejected during one EMRI orbit with probability of the form
\begin{align}
    p_{\rm ej} (\rpEMRI, \eEMRI, \rpDM, \eDM, \iota, q) = A(\rpEMRI, \eEMRI, \rpDM, \eDM, \iota) q^2\,,
\end{align}
where $A$ is a proportionality coefficient independent of $q$. We can now combine this with the number of orbits per EMRI and the total number of EMRIs an MBH experiences after 2.14 Gyr to obtain the cumulative survival probability of a single DM particle or, equivalently, the fraction of DM that remains.
Consider a single EMRI that completes $N_\mathrm{total}$ orbits over its lifetime, with periapsis $\rpEMRIi$ and eccentricity $\eEMRIi$ on its $i$th orbit in the loss cone. Assuming that the probability of survival over each EMRI orbit is independent of that of the previous orbit, the probability, $P_{\rm stay}$, that a given DM particle survives this EMRI is, to lowest order in $q$
\begin{align}
    \ln P_{\rm stay} (\rpDM, \eDM) \approx - q^2 \langle A \rangle_{{\rm orbs}} \,N_{\rm total}\,, \label{eq:ln P_stay EMRI lifetime}
\end{align}
where $\langle A \rangle_{\rm orbs} = \sum_i^{N_{\rm total}} A (\rpEMRIi, \eEMRIi, \rpDM, \eDM, \iota)/N_{\rm total}$. We validate the assumption of independent orbit survival probabilities against full-inspiral $N$-body simulations in App.~\ref{appsec:Validating Orbit-Independent Ejection Probabilities}. The evolution of $\rpEMRIi$ and $\eEMRIi$ from the initial conditions $(\rpEMRIz, \eEMRIz)$ is discussed in Appendix~\ref{appsec:Time Evolution of an EMRI}, where we show that the total number of orbits scales as
\begin{align}\label{eqn:N total with q}
    N_{\rm total} = \int_{f_{\rm init}}^{\infty} \frac{f}{\dot{f}}\, \dd f \equiv B(\rpEMRIz, \eEMRIz)\,q^{-1}\,,
\end{align}
for $q \ll 1$, where $f$ denotes the orbital frequency and $B(\rpEMRIz, \eEMRIz)$ is implicitly defined as the proportionality coefficient and is independent of $q$ for $q \ll 1$ (see Appendix \ref{appsec:Time Evolution of an EMRI}). 

We derive an `effective' $\langle A \rangle_{{\rm orbs}}$ when averaging over inclination, given by
\begin{align}
    \langle A \rangle_{{\rm orbs, \, eff}} = -\frac{1}{q^2 N_{\rm total}} \ln \left(\int P_{\rm stay} \, f_\iota \, \dd \iota\right) \label{eq:A_orb,eff}
\end{align}
where $f_\iota = \sin \iota/2$ ensures inclination is distributed uniformly in $\cos \iota$.

Averaging over the EMRI mass distribution, $f_q$ (which we take to consist of $\mEMRI=10\,\Msolar$ EMRIs only, so $f_q=\delta(q-10 \, \Msolar/\mSMBH)$), the EMRI periapsis and eccentricity distribution, $f_{\rpEMRIz}$, and assuming that each EMRI arrives with random orientation, the fraction, $f_{\rm rem}$, of DM that remains for a given periapsis and eccentricity is
\begin{align}
    \nonumber \ln f_{\rm rem} (\rpDM, \eDM) & = - N_{\rm EMRI} \int q \, f_q \, \dd q \int \dd \rpEMRIz \, f_{\rpEMRIz} \\ & 
    \quad \times \, B(\rpEMRIz, \eEMRIz) \, \langle A \rangle_{{\rm orbs, \, eff}} (\rpEMRIz, \eEMRIz, \rpDM, \eDM)\,, \label{eq:ln f_rem}
\end{align}
where $\eEMRIz = e_{\rm LC}(\rpEMRIz)$, $N_\mathrm{EMRI}$ is the total number of EMRIs. Eq.~\eqref{eq:ln f_rem} can be rewritten to show that, for a given MBH mass, $\ln f_{\rm rem} (\rpDM, \eDM) \propto N_{\rm EMRI} \langle \mEMRI \rangle$ where $\langle \mEMRI \rangle = \mSMBH \langle q \rangle$ is the average sBH EMRI mass with $\langle q \rangle = \int q \, f_q \, \dd q$. Rewriting this, we find $N_{\rm EMRI} \propto \ln f_{\rm rem} (\rpDM, \eDM)/\langle \mEMRI \rangle$ and thus, for a given $f_{\rm rem}$, the required EMRI rate is inversely proportional to the average EMRI mass.

We obtain an interpolated numerical expression for $\langle A \rangle_{{\rm orbs, \, eff}}$ using \multistar by performing a parameter scan in ($\rpEMRI, \eEMRI, \rpDM, \eDM$) space.  In particular, we scanned over $\eEMRI = \{$0, 0.2, 0.4, 0.6, 0.8, 1.0$\}$, $\eDM = \{$0.2, 2/3$\}$, $\rpDM/\RSMBH = \{$5, 6.5, 8.5, 10, 20, 50, 100, 200, 300, 500, 1000$\}$ and take 7 EMRI periapsis values $\rpEMRI = \rpDM + k(\raDM - \rpDM)$ for $k = \{0, 0.05, 0.25, 0.5, 0.75, 0.95, 1\}$.  For each of the resulting 924 grid points, we ran $10^6$ simulations of a single EMRI of mass ratio $q = 10^{-2}$ and a single DM particle over one EMRI orbit with each run sampling the initial EMRI and DM orbital parameters as described in \S\ref{subsec:Our 3b Model}.  Further, we set $A=0$ whenever $\rpEMRI < \rpDM$ or $\rpEMRI > \raDM$.  More strictly, the first condition should read $\raEMRI < \rpDM$; for computational efficiency, however, we neglect the intermediate regime $\rpEMRI < \rpDM < \raEMRI$.  These configurations are not expected to contribute significantly since the apoapsis of eccentric EMRIs decays rapidly during GW-induced orbital decay (Appendix~\ref{appsec:Time Evolution of an EMRI}).  Their inclusion, however, could only enhance DM ejection rates. 

\subsection{Numerical Results}
\label{subsec:EMRI Results}

We begin by computing the minimum EMRI rate required before $z = 3$ (2.14 Gyr) for the remaining fraction of DM particles to fall below $f_{\rm rem} = 10^{-6}$ for fixed DM periapsis and DM eccentricity using Eq.~\eqref{eq:ln f_rem}. Figure~\ref{fig:min_EMRI_rp_mS} shows this as a function of MBH mass, $\mSMBH$, and DM periapsis, $\rpDM$ for DM eccentricities $\eDM = 0.2$ (\textit{top panel}) -- chosen to illustrate the behaviour at low eccentricity -- and $\eDM = 2/3$ (\textit{bottom panel}), the average of a thermal eccentricity distribution. We take the MBH to be in the range $10^4 \Msolar < \mSMBH < 10^7 \Msolar$, as discussed in \S\ref{subsec:Param Distributions}. The upper boundary in each panel arises because the FP model predicts no EMRIs with such large periapsides and so no DM evaporation occurs \footnote{In that region the EMRI inspiral is slower than the FP diffusion due to stars. However, we would still expect the evaporation rate to be small, but non-zero, there due to the presence of BHs and neutron stars.}. The value of $f_{\rm rem} = 10^{-6}$ is an arbitrary choice used to demonstrate the parameter dependence of ejection probabilities; we will consider different ejection fractions below.
The required EMRI rates for given $f_{\rm rem}$ are slightly lower for $\eDM = 2/3$ than for $\eDM = 0.2$. This is because, for a given periapsis, DM particles with larger orbital eccentricities have larger apoapsides. At these larger radii, EMRIs inspiral more slowly and hence these DM particles are exposed to more ejection opportunities, reducing the required EMRI rate.
The maximum periapsis, $\rpLISA$, an EMRI can have such that its peak GW frequency in the LISA band is overlaid for reference. Taking the lower frequency threshold to be $f_{\rm LISA, min} = 10^{-4}$ Hz gives \citep[][Eq.~(36)]{Wen_2003}
\begin{align}
    \frac{\rpLISA}{\RSMBH} \approx 37 \left(\frac{1}{1+\eEMRI}\right)^{0.2} &\left(\frac{10^{-4} \, \mathrm{Hz}}{f_{\rm min}}\right)^{2/3} \times \nonumber\\
    & \left(\frac{10^6 \Msolar}{\mSMBH(1+z)}\right)^{2/3} \,,\label{eq:rp_LISA}
\end{align}
We see very weak eccentricity dependence and plot this for $\eEMRI = 0$ in Figure~\ref{fig:min_EMRI_rp_mS}.

\begin{figure}
    \centering
    {\includegraphics[width=0.48\textwidth]{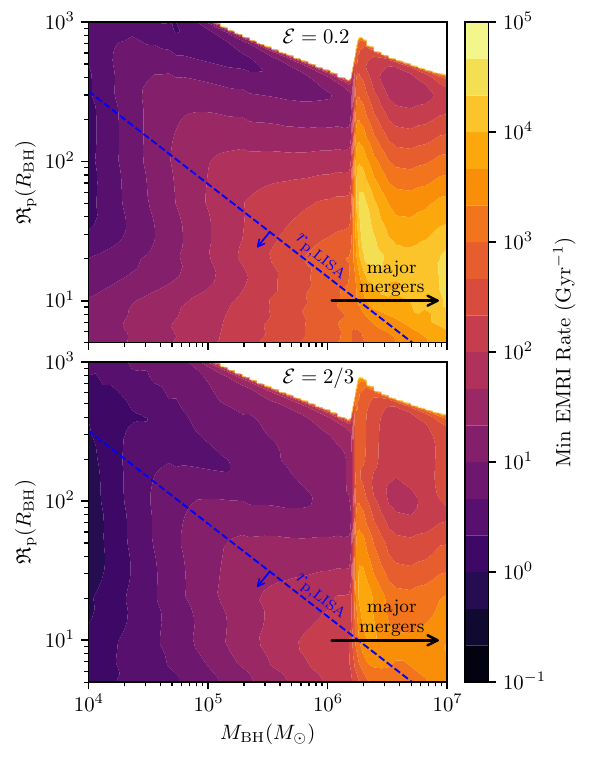}}
    \caption{Minimum EMRI rate required to reduce the fraction of remaining DM particles to below $f_{\rm rem} = 10^{-6}$ by $z = 3$ (2.14 Gyr) assuming all EMRIs have mass $\mEMRI = 10\Msolar$. We show this for $\eDM = 0.2$ (\textit{top panel}) and $\eDM = 2/3$ (\textit{bottom panel}). The blue arrow and dashed line mark $\rpLISA$ (Eq.~\ref{eq:rp_LISA}). This delineates the reduced DM density a typical LISA-type EMRI would encounter. The upper boundary (white region) comes from the fact that no EMRIs with such a large periapsis are predicted by FP and hence no evaporation is expected.  We also indicate the regime in which major mergers are likely to have already disrupted the DM spike. The change in behaviour for $\mSMBH \gtrsim 1.7 \times 10^6 \Msolar$ is owed to the relaxation time-scales moving above 2.14 Gyr.}
    \label{fig:min_EMRI_rp_mS}
\end{figure}
\begin{figure*}
    \centering
    {\includegraphics[width=0.99\textwidth]{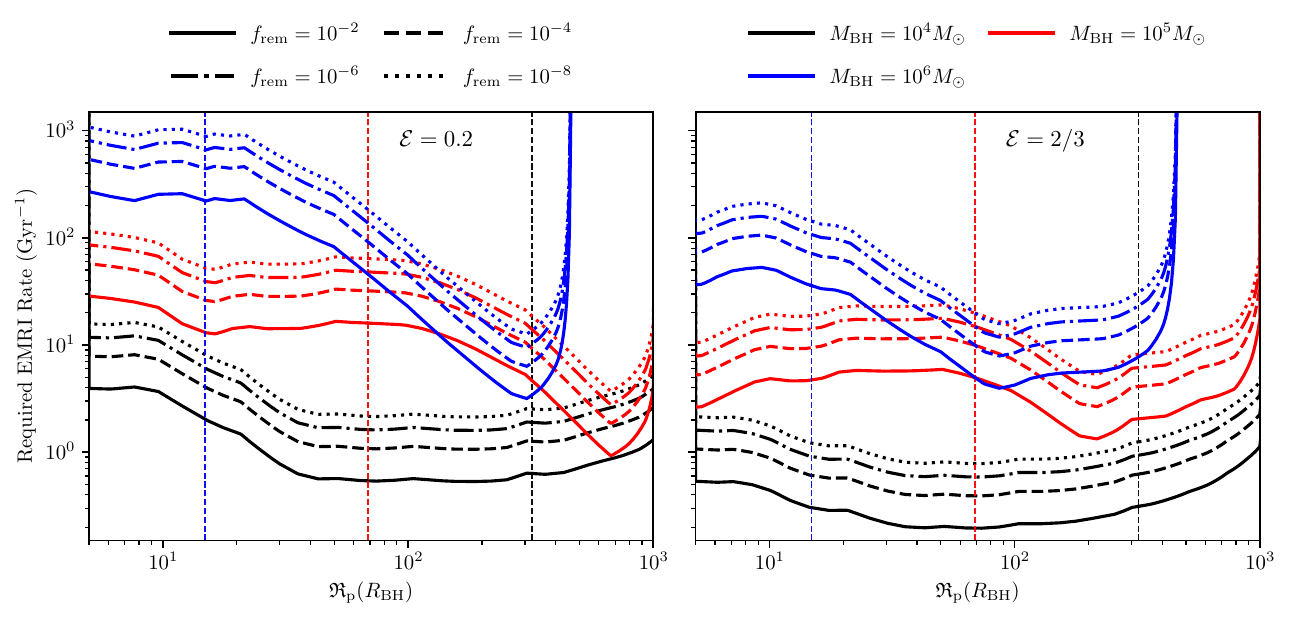}}
    \caption{EMRI rate required to deplete the DM spike for various MBH masses and ejection fractions by $z = 3$ for $\eDM = 0.2$ (\textit{left panel}) and $\eDM = 2/3$ (\textit{right panel}). We assume all EMRIs have mass $\mEMRI = 10\Msolar$. The vertical coloured lines indicate $\rpLISA$ (Eq.~\ref{eq:rp_LISA}) for each MBH mass.}
    \label{fig:min_EMRI_rates_f_rem}
\end{figure*}
The minimum EMRI rate in Figure~\ref{fig:min_EMRI_rp_mS} exhibits a non-trivial dependence on $\mSMBH$, which can be understood as follows: in our three-body models, we fix the mass of the sBHs contained in EMRIs to $\mEMRI=10 \Msolar$. Increasing $\mSMBH$ therefore decreases the mass ratio, $q = \mEMRI/\mSMBH$. From Eq.~\eqref{eq:ln f_rem}, $\ln f_{\rm rem}\propto -\langle q \rangle N_{\rm EMRI} $, thus a smaller mass ratio requires a larger EMRI rate for the same fraction of DM ejection, leading to the positive gradient toward larger MBH masses. However, there is a sharp (but continuous) change in behaviour for $\mSMBH \gtrsim \mUR \equiv 1.7 \times 10^6 \Msolar$ due to such systems being unable to relax within 2.14 Gyr, which significantly changes $f_{\rpEMRIz}$. This can be seen in Figure \ref{fig:f_rp_dists}, where the $\mSMBH = 2 \times 10^6 \Msolar$ distribution is heavily skewed towards smaller EMRI periapsides (see Appendix \ref{appsec:EMRI dis fn}).

In a similar vein, we can use Eq.~\eqref{eq:ln f_rem} to find the required EMRI rate, as a function of MBH mass, for various values of the fraction of the remaining DM, $f_{\rm rem}$. This is shown in Figure~\ref{fig:min_EMRI_rates_f_rem}, where the required EMRI rate around MBHs of mass $\mSMBH = 10^4 \Msolar$, $10^5 \Msolar$, $10^6 \Msolar$, and $10^7 \Msolar$ is shown for $f_{\rm rem} = 10^{-8}$, $10^{-6}$, $10^{-4}$, and $10^{-2}$. Vertical lines represent $\rpLISA$. We can clearly see that small changes in the EMRI rate lead to large changes in the depletion. This is because $\ln f_{\rm rem} \propto N_{\rm EMRI}$, as seen in Eq.~\eqref{eq:ln f_rem}, and thus $f_{\rm rem}$ depends exponentially on $N_{\rm EMRI}$. 
Eq.~\eqref{eq:ln f_rem} also implies that, for a given $f_{\rm rem}$, the required number of EMRIs scales as $N_{\rm EMRI} = \langle q \rangle ^{-1} |\ln f_{\rm rem}|\,  \langle B \langle A \rangle_{{\rm orbs, \, eff}} \rangle_{\rpEMRIz}^{-1} $. Here $\mEMRI = 10 \Msolar$ is fixed. However, the scaling with MBH mass is non-uniform and depends on the DM periapsis; this is a consequence of the mass dependence of $f_{\rpEMRIz}$. 

In particular, any EMRI whose initial periapsis exceeds that of a given DM particle will inevitably inspiral past that particle and can therefore strongly contribute to its evaporation probability. All EMRIs eventually pass through the innermost regions during the inspiral, whence the evaporation rate of DM particles at small periapsides is insensitive to the shape of the EMRI periapsis distribution at larger radii. Instead, the depletion there depends primarily on the total number of EMRIs rather than on their initial periapsis distribution. Thus, decreasing the MBH mass by an order of magnitude produces an approximately order-of-magnitude decrease in the required EMRI rate.
At larger DM periapsis distance, however, only the subset of EMRIs with sufficiently large initial periapsis will pass the DM particle as they spiral into MBH. Naturally the evaporation rate thus becomes sensitive to the shape of the EMRI periapsis distribution. As seen in Figure \ref{fig:f_rp_dists}, the distributions scale non-uniformly with mass, leading to large variations in the required EMRI rates that are seen at large DM periapsis compared to the inner regions.
Finally, another diagnostic that illustrates the DM density a typical LISA-like EMRI would encounter is to calculate the total fraction of DM that remains within a periapsis less than $\rpLISA$. This is given by
\begin{align}
    F_{\rm rem} (\eDM) = 
    \frac{4\pi \int_{5\RSMBH}^{\rpLISA} f_{\rm rem} (\rpDM, \eDM) \, \rpDM^2 \, \rho_{\rm DM} \, \dd \rpDM}{4\pi \int_{5\RSMBH}^{\rpLISA}  \rpDM^2 \, \rho_{\rm DM} \, \dd \rpDM} \label{eq:F_rem} \,,
\end{align}
where we only integrate down to $\rpDM = 5\RSMBH$ as this is the closest distance to the MBH that we consider -- smaller distances would require full GR calculations. Figure~\ref{fig:integrated_over_rp_horiz} shows $F_{\rm rem}$ as a function of MBH mass and EMRI rate after 2.14 Gyr. For EMRI rates within the accepted range of $1$--$1000$\,Gyr$^{-1}$ per galaxy (indicated by horizontal red lines), the DM spike is depleted by multiple orders of magnitude in a large portion of the MBH mass range considered. Again, we observe a strong change in behaviour for $\mSMBH \geq \mUR$.
\begin{figure*}
    \centering
    {\includegraphics[width=0.99\textwidth]{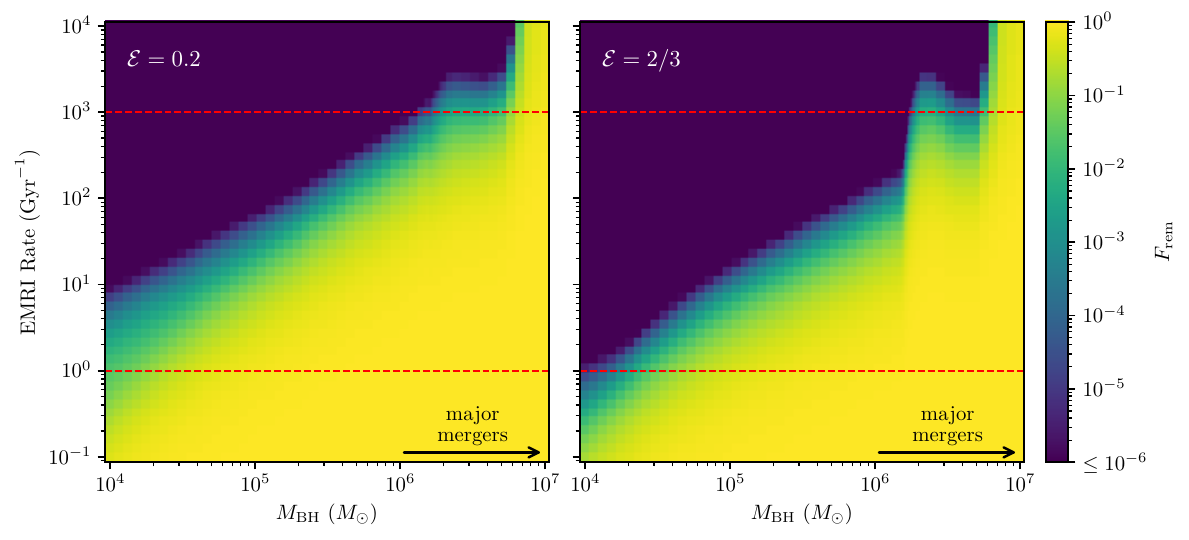}}
    \caption{Total fraction, $F_{\rm rem}$, of remaining DM integrated up to $\rpLISA$ (cf.~Eq.~\ref{eq:F_rem}) after 2.14 Gyr for $\eDM = 0.2$ (\textit{left panel}) and $\eDM = 2/3$ (\textit{right panel}), assuming an initial $\rho_{\rm DM} \propto r^{-7/3}$ spike and that all EMRIs have mass $\mEMRI = 10\Msolar$. The horizontal red lines indicate the commonly adopted lower and upper bounds on EMRI rates (see \S\ref{subsec:Param Distributions}). The black arrow denotes the parameter space where mergers are expected to have already disrupted the DM spike. Again, the change in behaviour for $\mSMBH \gtrsim 1.7 \times 10^6 \Msolar$ is owed to the relaxation time-scales moving above 2.14 Gyr, causing a qualitative change to the $(\rpEMRI,\eEMRI)$ distribution.}
    \label{fig:integrated_over_rp_horiz}
\end{figure*}
As mentioned in \S{\ref{subsec:Param Distributions}} above, EMRI rates are generally predicted to lie in the range $1$--$1000\,\Gyr^{-1}$ per galaxy, with many estimates clustering around $\mathcal{O}(100$--$300)\,\Gyr^{-1}$. Furthermore, as shown in Figures \ref{fig:min_EMRI_rp_mS}, \ref{fig:min_EMRI_rates_f_rem}, and \ref{fig:integrated_over_rp_horiz}, it is clear that the impact of EMRIs on DM spikes is highly sensitive to this rate. In the most optimistic scenario, with rates near $\sim 1000\,\Gyr^{-1}$, we expect nearly complete evaporation -- by many orders of magnitude -- of DM within the loss cone around all MBHs with masses $\mSMBH \lesssim 1.7 \times 10^6 \Msolar$. At the opposite extreme, even for a low rate of $\mathcal{O}(3-10)\,\Gyr^{-1}$ we still anticipate near-complete depletion for $\mSMBH \lesssim 10^5 \Msolar$. Recall, however, that masses above $10^{6}$--$10^{7} \Msolar$ would most likely have been through an equal-mass merger event in the past \citep{Conselice_2006, Fakhouri_2010, Ravi_2015, Rodriguez_Gomez_2016, OLeary_2021}, thereby disrupting the DM spike anyway. Thus, across a significant portion of the viable MBH mass range, EMRIs are expected to reduce the DM spike substantially.

\section{Discussion}
\label{sec:Discussion}

We now discuss the phenomenological consequences of our findings and their sensitivity to modelling choices.

\subsection{GW Dephasing Signatures}
\label{subsec:GW Dephasing Signatures}
A key phenomenological consequence of dense DM spikes is the GW dephasing they induce in IMRIs and EMRIs through effects such as accretion and dynamical friction. Due to the extraordinary sensitivity of next-generation detectors such as the Einstein Telescope or LISA, even phase shifts of order $\mathcal{O}(1)$ radian may be detectable \citep{Hinderer_2008}. Although existing estimates of DM-induced dephasing span several orders of magnitude, most works -- typically assuming steep density profiles with $\gamma_{\rm DM} \gtrsim 7/3$ -- predict cumulative phase shifts of $10^2$--$10^6$ radians over a multi-year observation \citep{Eda_2013, Yue_2018, Kavanagh_2020, Coogan_2022, Cole_2022, Dai_2022, Speeney_2022, Dosopoulou_2024, Montalvo_2024, Mukherjee_2024, Karydas_2024, Mitra_2025}.
To leading order, the accumulated dephasing scales linearly with the non-GR dissipative energy loss during the inspiral, which, in the case of dynamical friction and accretion, in turn scales linearly with the local DM density. Consequently, if repeated EMRIs reduce the density by a factor $f_{\rm rem}$, the dephasing is approximately suppressed by the same factor.
Applying this scaling to the upper end of existing predictions -- about $10^6$ radians over a 5-year LISA observation -- a depletion factor of $f_{\rm rem} \lesssim 10^{-6}$ would reduce the accumulated phase shift to less than $\mathcal{O}(1)$ radian. At this level, DM-induced effects may become indistinguishable from vacuum evolution, or at the very least, reduce the ability to recover DM parameters from the waveform. We stress that the precise detectability threshold depends on waveform systematics, observation time, signal-to-noise ratio, and related factors. For instance, Figure~3 in \citet{Cole_2022} indicates that if the DM density is reduced by $\gtrsim 2$ orders of magnitude, the SNR loss relative to the best-fit vacuum waveform drops to effectively zero for a one-year signal. Taken together, it is clear that the prospects for studying DM through LISA are reduced by the results presented here. Aside from fine-tuned scenarios, this substantially narrows the mass range over which collisionless DM spikes could be detected with future GW detectors: for masses between $10^4 \, \Msolar$ and ($10^6$--$10^7$)\,$M_{\odot}$, our results show these spikes are severely depleted by $z=3$, while for heavier BH masses, the spikes are likely to have already been destroyed by major mergers. Thus detecting the presence of DM spikes through GW dephasing likely requires detecting EMRIs with very high signal-to-noise ratios at much higher redshifts than $z=3$.
The situation may be qualitatively different for self-interacting DM (SIDM), where elastic scattering can redistribute energy and angular momentum. The relevant time-scale here is the collisional relaxation time,
\begin{align}
    t_{\rm relax} \sim \frac{1}{\rho_{\rm DM}\bar{\sigma}\mathcal{V}}\,,
\end{align}
where $\bar{\sigma}$ is the cross section per unit mass and $\mathcal{V}$ the DM particle's velocity. For typical spike densities and cross sections near current upper limits ($\bar{\sigma} \lesssim 0.1$--$1\,\mathrm{cm^2/g}$ \citep{Tulin_2018,Eckert_2022,Adhikari_2024}), this relaxation time can be relatively short, especially in the inner regions. Collisional transport could thus efficiently refill the phase space depleted by EMRIs, partially replenishing the spike. This may therefore be an effective mechanism for stabilising the spike against EMRI-induced depletion and preserving detectable GW dephasing. 
Meanwhile, the stellar sphere is observationally challenging to probe: DM densities there are already reduced relative to the innermost region due to the steep density gradient, and GWs emitted by sources at such distances are difficult to detect. 
Shorter relaxation times would strengthen these imprints by more efficiently resupplying the DM that is removed from the system during the binary's inspiral. Further, it could influence the formation of DM mini-spikes around sBHs located in the stellar sphere and galactic field that could be detectable with \emph{Gaia} \citep{Branco_2025} and could affect the dissipation of orbital energy by BHs moving through the stellar sphere.
\subsection{DM Annihilation Signatures}
\label{subsec:DM Annihilation Signatures}
Another phenomenological consequence of DM over-densities near MBHs is the enhanced high-energy particle flux in self-annihilating DM (SADM) models. In such scenarios, the predicted annihilation flux scales as $\rho_{\rm DM}^2$ \citep{Lee_1977}, so any suppression of the central density leads to a substantial reduction in the expected signal. Numerous studies have explored the potential of current and future $\gamma$-ray observations to constrain annihilating DM in the presence of density spikes \citep{Bergstrom_2012, Alvarez_2021, Freese_2022}. Others argue that existing $\gamma$-ray observations already suggest either that DM is not self-annihilating, or that extremely dense DM over-densities do not exist \citep[][and references therein]{Bertone_2006, Aharonian_2008, Bertone_2009, Wanders_2015, Balaji_2023, Bertone_2024}. The spike evaporation channel identified here supports the latter interpretation:~EMRI-induced heating can significantly weaken indirect-detection bounds that assume long-lived, collisionless spikes made of cold DM.
If DM is self-annihilating, our conclusions regarding spike density are only strengthened. The characteristic annihilation time-scale,
\begin{align}
    t_{\rm ann} \sim \frac{m_{\rm DM}}{\rho_{\rm DM}\langle \sigma v\rangle}\,,
\end{align}
implies the existence of a critical density above which annihilation efficiently depletes the spikes over cosmological time-scales. For typical WIMP parameters ($m_{\rm DM} \sim 100\,\mathrm{GeV}$ and $\langle \sigma v \rangle \simeq 3 \times 10^{-26}\,\mathrm{cm}^3\,\mathrm{s}^{-1}$ \citep{Steigman_2012}), requiring $t_{\rm ann} \sim \mathrm{Gyr}$ yields
\begin{align}
\rho_{\rm crit} \sim 1.1 \times 10^{11} \, \frac{\rm GeV}{\mathrm{cm}^3} \approx 3 \times 10^9 \frac{\Msolar}{\pc^3} \,.
\end{align}
Densities exceeding this value are rapidly depleted to $\sim \rho_{\rm crit}$ on Gyr time-scales. Annihilation therefore sets an upper bound on the long-term spike density. Combined with EMRI-driven depletion, the quadratic scaling of the flux renders the expected annihilation signal small. In this sense, our collisionless treatment likely overestimates the surviving densities in SADM scenarios.

\subsection{Sensitivity to Model Assumptions}
\label{subsec:Sensitivity to Model Assumptions}

We review the main parameter choices and assumptions underlying our model. Where possible, we have made conservative choices, except for a few neglected processes that are justified throughout the paper and summarised below.
\begin{itemize}
    \item We assume that the initial $r^{-7/3}$ collisionless spike is smooth, isotropic, and phase mixed, with a thermal eccentricity distribution $f_{\eDM} = 2\eDM$. A strongly clumped or globally anisotropic DM distribution is beyond the scope of this work.

    \item The EMRI mass distribution influences DM ejection fractions strongly. In particular, Eq.~\eqref{eq:ln f_rem} implies
    \begin{align}\label{eq:ln mathcal_f_rem_summary}
         \ \ \ \ln f_{\rm rem} \propto N_{\rm EMRI}\int q f_q \dd q = \frac{N_{\rm EMRI} \langle \mEMRI \rangle}{\mSMBH}\,,
    \end{align} 
    where $q=\mEMRI/\mSMBH$ so that the fraction of remaining DM depends exponentially on the accumulated EMRI mass during the prior evolution of the MBH, for a fixed $\mSMBH$. A heavier EMRI population would enhance DM evaporation substantially. However, since the mass distribution of EMRIs is not well constrained and is sometimes assumed to be dominated by sBHs of mass 10--30$\, \Msolar$ or heavier \citep{Barack_2004, Hopman_2009, Aharon_2016, Babak_2017}, we made the conservative choice of $\langle\mEMRI\rangle = 10 \, \Msolar$.
    \item We treat the EMRI rate as a free parameter. Although most studies favour values of $\mathcal{O}(100 - 300)\,\Gyr^{-1}$ per galaxy (with estimates typically spanning the range $1$--$1000\,\Gyr^{-1}$ ; see \S\ref{subsec:Param Distributions}), we explore a broader interval in order to assess the full range of possible ejection fractions (see \S\ref{subsec:EMRI Results} and \S\ref{subsec:GW Dephasing Signatures}).
    \item The EMRI rate required to deplete the DM spike depends on the redshift of the observed EMRI, as this sets the accumulated EMRI mass by the age of the MBH in Eq.~\eqref{eq:ln mathcal_f_rem_summary}.
    Since $f_{\rm rem}$ depends exponentially on $N_{\rm EMRI}\langle \mEMRI\rangle$, lower-redshift systems --- with longer available evolution times --- will have been depleted more and hence produce weaker dephasing signatures, while higher-redshift systems yield stronger dephasing. We conservatively used the age corresponding to the upper redshift limit for EMRI detections by LISA, $z = 3$ \citep{Babak_2017}.
    \item We have neglected the influence of direct plunge EMRI orbits on the DM distribution \footnote{A related, and often more constraining, assumption was to neglect the effects of EMRIs with $\rpEMRI < \rpDM < \raEMRI$ for a given DM periapsis, $\rpDM$.}.  While these are expected to contribute only weakly to EMRI observation rates --- due to their small periapsis and rapid merging times (see \S{\ref{subsec:Param Distributions}}) --- they could nevertheless contribute to DM spike evaporation. However, given the large uncertainties in their rates and periapsis and eccentricity distributions, and the difficulty of modelling them, we omit them from our analysis. Including them would only make our conclusions on the inexorable DM depletion stronger.
    \item Our results have assumed a Hernquist initial density profile which is typically used to describe NSCs. However, different NSC formation models predict different initial density profiles (see Appendix \ref{appsec:Diff NSC ICs}). This can lead to a different critical mass, $\mUR$, below (above) which cusps have (have not) relaxed before $z=3$ and hence have (have not) ``forgotten'' their initial conditions. As a result, the critical EMRI rates above $\mUR$ exhibit dependence on the choice of initial profile. However, higher mass MBHs will typically have undergone major mergers which would have erased the DM spike independently of EMRIs for unrelaxed systems.
    \item We have assumed that the DM ejection probability per EMRI orbit is independent of previous orbits. While this manuscript was under review, \citet{Karydas_2026} appeared on the arXiv. They study the same physical system but take a very different approach to calculating the depletion. They argue that our assumption of independent ejection probabilities per orbit causes us to over-predict DM depletion. However, we believe this is not the case and that our results are consistent with theirs. We detail our arguments in Appendix \ref{appsec:Validating Orbit-Independent Ejection Probabilities}.
\end{itemize}
Several additional caveats remain: we have neglected the effects of strong encounters, resonant relaxation and relativistic apsidal precession within the SoI in our FP models and neglected resonant relaxation and Kozai-Lidov type effects within the loss cone when modelling the interaction of DM with EMRIs. While we expect these to be minimal (see \S\ref{subsec:FP and SSE}), modelling these processes is beyond the scope of this work. We also neglect the role of stellar evolution --- such as ongoing star formation and mass loss --- which can modify two-body relaxation in FP models \citep{Aharon_2015, Vasiliev_2017, Wang_2023}. We have assumed a static EMRI distribution taken to be that at $z = 3$. A more detailed analysis would assume an evolving EMRI distribution; however this would significantly increase the complexity of evaluating Eq.~\eqref{eq:ln f_rem}, and would most likely merely induce a slightly smoother transition across $\mUR$. Finally, and perhaps most importantly, we have assumed that the galactic environment of MBHs contains a NSC. Their presence is not guaranteed, particularly for higher mass MBHs \citep[e.g.,][and references therein]{Hoyer_2021,Neumayer_2020}.

\section{Conclusions}
\label{sec:Conclusions}

We have modelled the DM spike as a dynamical component of a realistic multi-mass galactic nucleus. We have identified two independent depletion mechanisms that significantly and irreversibly evaporate DM spikes on distinct spatial scales, particularly for MBHs with masses $\mSMBH \lesssim 10^6 \, \Msolar$. As a result, a canonical Gondolo--Silk spike does not represent a long-lived configuration of such systems.
We found that mass segregation in multi-mass stellar cusps shortens the DM relaxation time by a factor of $\mathcal{O}$(10--100) relative to single-effective-mass models, causing steep initial DM spikes to evolve towards the much lower-density Bahcall--Wolf profile $\rho \propto r^{-3/2}$ much earlier than previously assumed. We use these multi-mass models to set the initial conditions for the EMRIs which deplete the DM spike in the inner region. 
%

%
In the loss cone, strong encounters between dark matter particles and stellar-mass black hole EMRIs eject dark matter particles through gravitational slingshot interactions, progressively depleting the spike over Gyr time-scales. We focussed on the portion of the dark matter spike enclosed within $\rpLISA$, the maximum periapsis at which an EMRI can enter the LISA band (Eq.~\ref{eq:rp_LISA}). The most optimistic existing predictions of spike-induced dephasing reach $\sim 10^6$ radians; a density reduction by a comparable factor would suppress even these extreme dephasings to $\mathcal{O}(1)$ radian -- the sensitivity threshold for LISA. Our results indicate that even conservative EMRI rates of $\sim \mathcal{O}(3 - 10) \, \Gyr^{-1}$ per galaxy are sufficient for MBHs in the range $10^4 \Msolar \leq \mSMBH < 10^5 \Msolar$ to undergo such a depletion before $z=3$. For more typical rates of $\sim$100--300$\, \Gyr^{-1}$, this extends up to $\mSMBH \lesssim 1.7 \times 10^6 \Msolar$, above which our results are strongly dependent on the star cluster's initial conditions -- which are poorly constrained -- due to the EMRI energy distribution not having relaxed by $z=3$. Further, these required EMRI rates are inversely proportional to the average EMRI mass, which we took to be $10 \Msolar$.

We emphasise that these results are based on conservative assumptions of EMRI masses and the cosmological redshift of EMRIs observed by LISA. Due to the ejection fractions being exponentially sensitive to these parameters, our results would be strengthened by more realistic estimates.
EMRI-driven evaporation therefore substantially narrows the parameter space where the collisionless $r^{-7/3}$ Gondolo--Silk-type spikes can plausibly survive. This has direct implications for LISA: the MBH mass window once considered most promising for detectable DM spike signatures might, in practice, be largely discarded due to EMRI-driven evolution alone. 
Dense DM spikes could, in principle, still survive at very high redshifts of $z\gg 3$ or in a limited range of masses and EMRI rates. For example, around heavy MBHs with masses $\mSMBH \gtrsim 10^6 \, \Msolar$ that have not undergone any major merger with another MBH and with low EMRI rates, or in models where DM is self-interacting, in which depleted regions can be partially replenished through collisions, thereby resurrecting the spike. Additionally, our results do not exclude the possibility that lighter, adiabatically formed IMBHs ($\mSMBH \leq 10^4 \, \Msolar$), or MBHs without nuclear clusters, retain DM spikes.
Our results have demonstrated that the influence of realistic galactic environments must be incorporated in any assessment of DM-induced GW signatures. Ignoring this evolution risks significantly overestimating the prospects for detecting DM spikes with LISA.

\section*{Acknowledgements}

We would like to thank John Magorrian for insightful conversations about strong encounters and direct plunge orbits, Fabio Antonini, Fani Dosopoulou and Johan Samsing for helpful discussions on DM spikes and EMRIs, and Ygal Klein for pointing out some references. This work was supported by the STFC (grant No.~ST/W000903/1), and by a Leverhulme Trust International Professorship Grant (No.~LIP-2020-014). C.S.\ acknowledges funding from an STFC studentship. Y.B.G.'s work was partly supported by the Simons Foundation via a Simons Investigator Award to A.A.~Schekochihin. T.F.M.S.\ would like to thank the Institute for Advanced Study in Princeton for its hospitality during the final stages of this work. T.F.M.S.\ acknowledges support from a Royal Society University Research Fellowship (URF-R1-231065).  A.H.\ acknowledges support from the Australian Research Council through grants DP240101786 and DP240103174.

\section*{Data Availability}

Data will be made available upon reasonable request to the corresponding author. \phaseflow, which is part of the \agama package, is publicly available \citep{Vasiliev_2018}\footnote{\url{https://eugvas.net/software/phaseflow/}}; \multistar documentation is available online \footnote{ \url{https://2sn.erc.monash.edu/multistar/doc/index.html}.}.



\bibliographystyle{apsrev4-2}
\bibliography{apssamp} 



\appendix

\section{Different Initial NSC Density Profiles}
\label{appsec:Diff NSC ICs}

In \S\ref{sec:DM Spikes in Mass Segregated Stellar Cusps}, we assumed that our NSC initially obeyed a Hernquist density profile (Eq.~\eqref{eq:Hern stellar profile}), which scales as $r^{-1}$ inside the sphere of influence. However, different NSC formation histories predict different profiles. Formation via in situ star formation predicts inner profiles with $\rho_* \propto r^{- \gamma_{{\rm star},i}}$ with $\gamma_{{\rm star},i} \in [2.4,2.95]$ \citep{Bartko_2009, Bartko_2010}, while formation via the inspiral of globular clusters predicts $\gamma_{{\rm star},i} = 0.5$ \citep{Antonini_2012,Antonini_2013}. Different slopes lead to different relaxation times. To examine this dependence, we modify our density profile by generalising the Hernquist profile to the Dehnen profile \citep{Dehnen_1993}
\begin{align}
    \rho(r)=\rho_0\left(\frac{r}{a}\right)^{-\gamma}\left[1+\frac{r}{a}\right]^{\gamma-4} \,.
\end{align}
While Hernquist corresponds to $\gamma = 1$, we now take $\gamma = \gamma_{{\rm star},i}$. Figure \ref{fig:relax_diff_cs} shows the relaxation time-scales at $t = 2.14 \, \Gyr$ for various values of $\gamma_{{\rm star},i}$ in our extended-mass-function model (\textit{solid} lines) and our $7.55 \, \Msolar$ single-effective-mass model (\textit{dashed} lines). We have selected $2.14 \, \Gyr$ as this corresponds to a redshift of $z = 3$ --- the highest EMRI redshift that LISA can detect (see \S\ref{sec:DM and EMRIs}).

In all cases, the relaxation time-scales in our multi-mass model are shorter than the single-mass model, which in turn leads to more rapid reductions in DM densities. Thus, even for different initial conditions, shorter relaxation times are a generic consequence of using multi-mass models in comparison to single-effective-mass models. This strengthens our conclusions in \S\ref{sec:DM Spikes in Mass Segregated Stellar Cusps}.

\begin{figure}
    \centering
    \includegraphics[width=0.49\textwidth]{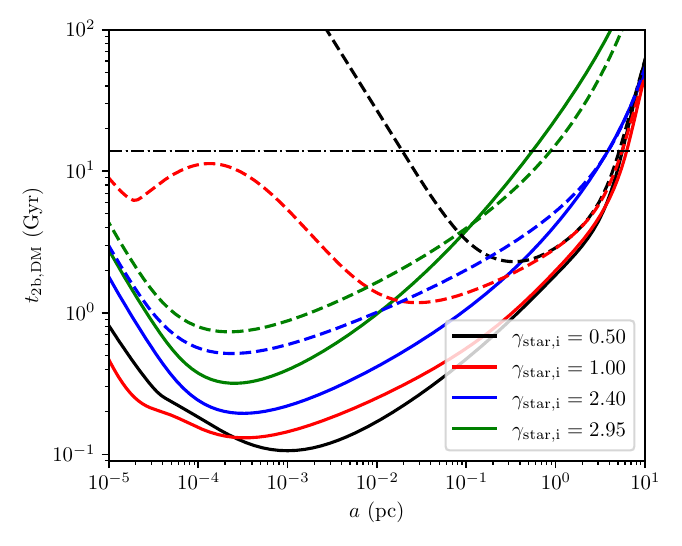}
     \caption{The relaxation time-scale after $t = 2.14 \, \Gyr$ for $\mSMBH = M_{\rm Sgr \, A^*} = 4.3 \times 10^6 \, \Msolar$ with different initial NSC density profile power-laws $\gamma_{{\rm star},i}$. The \textit{solid} lines represent our extended-mass-function model, while the \textit{dashed} lines represent the single-effective-mass ($7.55\, \Msolar$) model.
     }
    \label{fig:relax_diff_cs}
\end{figure}

However, this also has implications for the EMRI distribution which is a crucial parameter in \S\ref{sec:DM and EMRIs}. In particular, the critical MBH mass, $\mUR$, above which our NSC is no longer able to relax within $2.14 \, \Gyr$ depends on the initial density profile. For Hernquist, this was $\mUR \approx 1.7 \times 10^6 \, \Msolar$, as discussed in \S\ref{subsec:Param Distributions}. For shallower (steeper) profiles, this transition mass becomes smaller (larger). For $\mSMBH < \mUR$, the initial density profile is `forgotten' due to relaxation and thus our results are unchanged. For $\mSMBH > \mUR$, this is not the case, and the EMRI distribution will depend on the initial profile. Exploring this in more detail is left for future work.

\section{Evolution with the Truncated Mass Function}
\label{appsec:Evolution with the Truncated Mass Function}

%
\begin{figure*}
    \centering
        \subcaptionbox{Multi-mass stellar model with ten species logarithmically spaced between $0.1\,\Msolar$ and $10\,\Msolar$.
        \label{subfig:DM_sim10_threepanel_light}}
        {\includegraphics[width=0.99\linewidth]{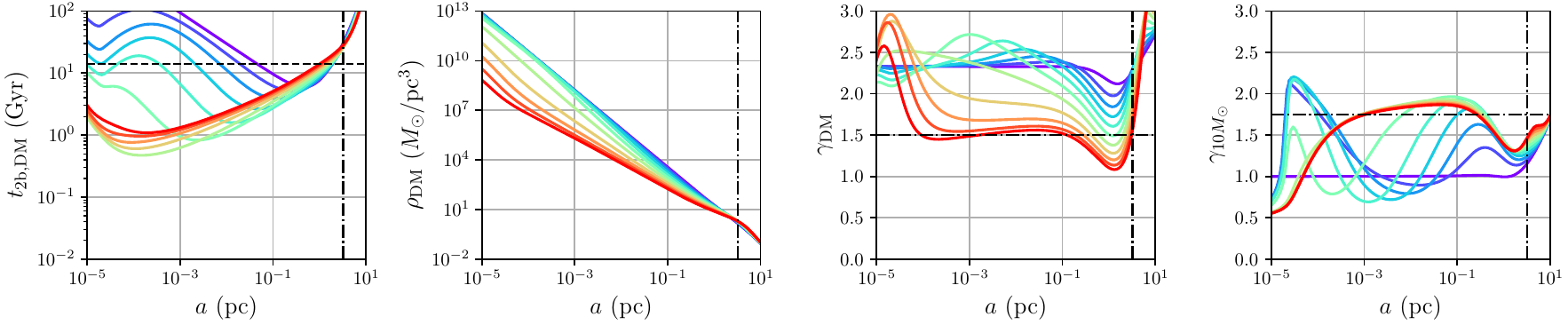}}\\\vspace{0.25cm}
        \subcaptionbox{Single-mass stellar model with mass $1.6\,\Msolar$.
        \label{subfig:DM_sim1_threepanel_light}}
        {\includegraphics[width=0.99\linewidth]{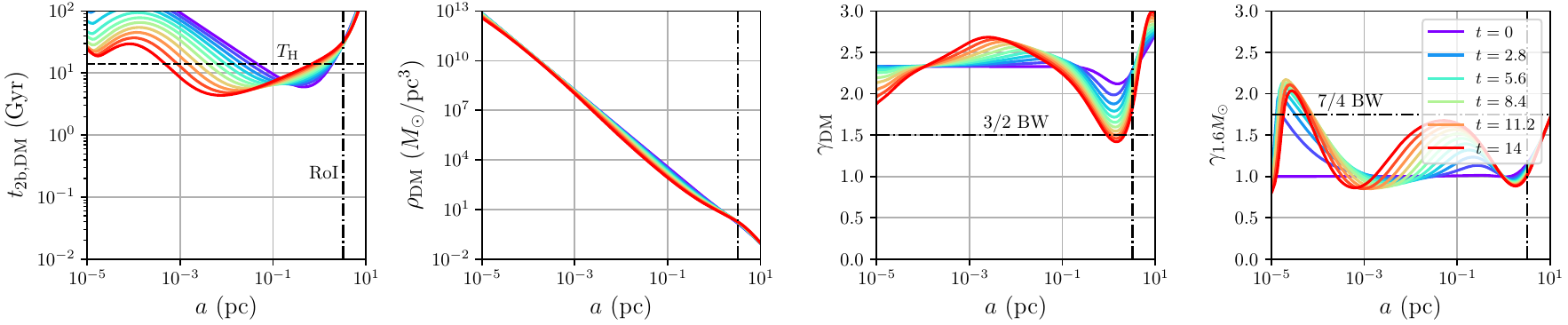}}
        \caption{
        Same as Figure~\ref{fig:DM_threepanel_plots_heavy}, but for the truncated mass function (see \S{\ref{subsec:SMF}}), and evolved for 14 Gyr. The multi-mass model again exhibits significantly faster relaxation.
        }
    \label{fig:DM_threepanel_plots_light}
\end{figure*}

We use \phaseflow to numerically integrate the FP equation (Eq.~\ref{eq:FP}) for our truncated mass function, where we take 10 logarithmically spaced stellar masses ranging from $0.1\,\Msolar$ to $10\,\Msolar$. In particular, we wish to study the effect of mass segregation on the relaxation times of such a system in comparison to one with a single stellar species of effective mass $1.6 \, \Msolar$ (Eq.~\ref{eq:eff_mass}). See \S\ref{subsec:Results -- FP} for further details. We show these results in Figure~\ref{fig:DM_threepanel_plots_light}. The \textit{top row} corresponds to the multi-mass model, and the \textit{bottom row} to the single-effective-mass model. From left to right, the panels display: the evolution of the DM two-body relaxation time-scale (Eq.~\ref{eq:t_2b rel}), the DM density profile, the DM density power-law slope (initially $\rho_{\rm DM} \propto r^{-7/3}$), and the density slope of the most massive stellar species (initially Hernquist; Eq.~\ref{eq:Hern stellar profile}). Because of the longer relaxation times compared to the extended mass function (due to the truncated model lacking the most massive stars included in the extended model), the system is evolved for 14 Gyr. The qualitative behaviour is the same as in the extended case (Figure~\ref{fig:DM_threepanel_plots_heavy}; in which masses range from $0.1\,\Msolar$ to $100\,\Msolar$) but proceeds on longer time-scales. Mass segregation enhances the central stellar density rapidly, thereby reducing relaxation times and accelerating the transition of the DM cusp towards the $\rho_{\rm DM} \propto r^{-3/2}$ BW profile.

\section{Details of Three-Body Simulations}
\label{appsec:Details of Three-Body Simulations}

In this appendix, we provide some details on the implementation of three-body simulations for \S \ref{sec:DM and EMRIs}, using \multistar. 

\subsection{Initial Conditions for each Relativistic Orbit}
\label{subsec:Rel Orbit ICs}

In Newtonian gravity, an orbit is fully specified by its semi-major axis, eccentricity, and orbital phase.  In GR, due to relativistic apsidal advance, bound orbits are not closed loops, and therefore a different approach is needed.  A convenient and intuitive alternative is to parametrise bound geodesic orbits through their radial turning points.
We denote the inner and outer turning points by $\rin$ and $\rout$, corresponding to periapsis and apoapsis, respectively. These uniquely determine the shape of the orbit (up to orientation) and can be used to obtain the specific energy, $E$, and the specific angular momentum, $L$, which then provide the time evolution of the particle’s position and velocity along the orbit through the Schwarzschild geodesic equations. The orbit can then be rotated to match any desired orientation.

%
Assuming equatorial motion ($\theta=\pi/2$) without loss of generality, the Schwarzschild geodesic equations are
\begin{subequations} \label{eq:Sch geodesic eqns}
    \begin{align}
        \left(\frac{\dd r}{\dd \tau}\right)^2 &= \frac{E^2}{c^2} - \mathrm{c}^2\left(1 - \frac{\RSMBH}{r}\right)\left(1 + \frac{L^2}{\mathrm{c}^2r^2}\right)\,,\\
        \frac{\dd \phi}{\dd \tau} &= \frac{L}{r^2}\,,\\
        \frac{\dd t}{\dd \tau} &= \frac{E}{\mathrm{c}^2\left(1 - \RSMBH/r\right)}\,.
    \end{align}
\end{subequations}
Enforcing $\dd r/\dd\tau=0$ at $\rin$ and $\rout$ yields
\begin{align}
    E^2 &= \frac{\mathrm{c}^4(\rin - \RSMBH)(\rout - \RSMBH)(\rin + \rout)}{\rin \rout (\rin + \rout) - \RSMBH(\rin^2 + \rin \rout + \rout^2)}\,,\\
    L^2 &= \frac{\mathrm{c}^2 \RSMBH \rin^2 \rout^2}{\rin \rout (\rin + \rout) - \RSMBH(\rin^2 + \rin \rout + \rout^2)}\,,
\end{align}
where we take $E, L > 0$ such that $\dd \phi/\dd \tau > 0$ and $\dd t/\dd \tau > 0$. The (cubic) equation for the radial turning points admits three roots; the physical orbital turning points correspond to the outer two real roots. To select these, we require
\begin{align}
    \rin \rout > \RSMBH (\rin + \rout + \max[\rin,\rout])\,.
\end{align}
The evolution of position and velocity then follows from Eq.~\eqref{eq:Sch geodesic eqns} through $v_r = \dd r/\dd t$ and $v_\phi = r \dd \phi/\dd t$. Note that the velocity is defined with respect to the coordinate $t$ rather than the proper time $\tau$, since we are interested in quantities as measured by an observer at infinity. 
We initialise the EMRI at apoapsis and randomise the DM particle’s orbital phase.  Since phase is not uniquely defined in GR, we generate the orbit numerically, sample uniformly in observer time, and select a random timestamp to define the initial position and velocity. 
A particle is classified as ejected if, at the end of the integration, its specific energy satisfies $E > \mathrm{c}^2$ and its radial velocity is positive.  Rearranging Eq.~\eqref{eq:Sch geodesic eqns} yields
\begin{align}
    E^2 = \frac{\alpha^3 \mathrm{c}^6}{-v_r^2 + \alpha(\alpha \mathrm{c}^2 - v_\phi^2)}\,,
\end{align}
where $\alpha = 1 - \RSMBH/r$.
These quantities must, however, be transformed from Schwarzschild coordinates to harmonic coordinates before being passed to the numerical code.

\subsection{Moving between Harmonic Coordinates and Schwarzschild Coordinates}

Care must be taken with the gauge choice when transforming between harmonic coordinates, $x_{\rm H}^\mu$ --- those used by \multistar \xspace --- and Schwarzschild coordinates, $x_{\rm sch}^\mu$ --- those used in \S\ref{subsec:Rel Orbit ICs}. The gauge chosen within the PN formalism ultimately determines the coordinate system of the code. \multistar uses the harmonic gauge, in which one uses harmonic coordinates $x_{\rm H}^\mu$ defined such that $\Box x_{\rm H}^\mu = 0$. We wish to find $x_{\rm H}^\mu (x_{\rm sch}^\mu)$. Since $\Box x_{\rm H}^\mu = \partial_\nu \left[(\sqrt{-g_{\rm sch}} g_{\rm sch}^{\nu\lambda}) \partial_\lambda x_{\rm H}^\mu\right]/\sqrt{-g_{\rm sch}}$, where $g_{\rm sch}$ is the metric in Schwarzschild coordinates and derivatives are with respect to $x_{\rm sch}^\mu$, then $x_{\rm H}^\mu$ must satisfy $\partial_\nu \left[(\sqrt{-g_{\rm sch}} g_{\rm sch}^{\nu\lambda}) \partial_\lambda x_{\rm H}^\mu\right] = 0$. Given that the Schwarzschild metric is stationary and spherically symmetric, we assume the coordinates are of the form $x_{\rm sch}^i = r \, n^i(\theta, \phi)$ and $x_{\rm H}^i = \rho(r) n^i(\theta, \phi)$, where $n^i$ are the usual spherical coordinate unit vectors. This yields a differential equation for $\rho(r)$
\begin{align}
    (r^2 - \RSMBH r) \rho''(r) + (2r - \RSMBH) \rho'(r) - 2\rho(r) = 0\,. \label{eq:ODE for rho(r)}
\end{align}
Defining $z = 2(r - \RSMBH/2)/\RSMBH$, one finds
\begin{align}
    (z^2 - 1) \frac{\dd^2 \rho}{\dd z^2} - 2z \frac{\dd \rho}{\dd z} + 2\rho = 0\,.
\end{align}
This is a Legendre differential equation with $\ell = 1$, whose solutions are known. Substituting these back into $z(r)$ gives us two independent solutions to Eq.~\eqref{eq:ODE for rho(r)}: $\rho(r) = r - \RSMBH/2$ and $\rho(r) = (r - \RSMBH/2) \ln((r - \RSMBH)/r) + \RSMBH$. The latter solution is singular at the MBH horizon, leaving a single physical solution, $\rho(r) = r - \RSMBH/2$. 
We hence find that
\begin{align}
    \V{x}_{\rm H} &= \left(1 - \frac{\RSMBH}{2 r}\right) \V{x}_{\rm sch} \,,\\
    \V{v}_{\rm H} &= \left(1 - \frac{\RSMBH}{2 r}\right) \V{v}_{\rm sch} + \frac{\RSMBH}{2 r^3} \left(\V{x}_{\rm sch} \cdot \V{v}_{\rm sch}\right) \V{x}_{\rm sch} \,,
\end{align}
where the latter is derived by taking the time derivative of the former. These can be inverted to get
\begin{align}
    \V{x}_{\rm sch} &= \left(1 + \frac{\RSMBH}{2 \rho}\right) \V{x}_{\rm H} \,,\\
    \V{v}_{\rm sch} &= \left(1 + \frac{\RSMBH}{2 \rho}\right) \V{v}_{\rm H} - \frac{\RSMBH}{2 \rho^3} \left(\V{x}_{\rm H} \cdot \V{v}_{\rm H}\right) \V{x}_{\rm H} \,.
\end{align}

\section{Mass Dependence of the EMRI Distribution Function}
\label{appsec:EMRI dis fn}

Here, we explain the subtleties behind the complicated mass dependence of the EMRI periapsis distribution shown in Figure \ref{fig:f_rp_dists}. In particular, we will discuss why the upper periapsis cut-off moves inwards as mass increases up to $\sim 10^6 \, \Msolar$, then suddenly jumps outwards before continuing to move inwards again. Secondly, we will give a qualitative explanation as to why relaxed systems have a greater contribution from EMRIs with large periapsis than unrelaxed ones, particularly for the $\mSMBH = 2 \times 10^6 \, \Msolar$ case.

The former is most easily explained by examining how the ratio of the angular momentum relaxation time-scale to the GW inspiral time-scale --- $T_J/T_{\rm GW}$ --- scales with EMRI eccentricity, EMRI periapsis, and MBH mass. The outer loss-cone boundary is defined as the region where this ratio is unity \footnote{Here we are discussing the outer loss-cone boundary in periapsis--eccentricity space, as opposed to the right boundary of Figure \ref{fig:f_rp_dists}.}. The relaxation time-scale is given by $T_J = (1 - \eEMRI^2) T_{\rm rel}$ where $T_{\rm rel} = \min[T_{\rm SRR}, T_{\rm 2b}]$ is dominated by $T_{\rm 2b}$ (SRR is quenched by relativistic apsidal precession; cf.~\S\ref{subsubsec:FPE}). Now, we know that $T_{\rm 2b} \propto \sigma^3/\langle \rho(a) \rangle$ (Eq.~\ref{eq:t_2b rel}) where $\langle \rho(a) \rangle$ is the stellar-mass-weighted average density and $a$ is the semi-major axis; for the sake of the argument, let us na\"{i}vely assume that this average density obeys some power-law scaling with $a$: $\langle \rho(a) \rangle \propto a^{-\delta}$. From Eqs.~(\ref{eq:Hern stellar profile}--\ref{eq:M_BH norm}), one finds that the normalisation constant of the stellar density is $\rho_{0,*} \propto \mSMBH^{0.048}$ --- that is, $\rho_{0,*}$ is effectively independent of mass. Additionally, the velocity dispersion satisfies $\sigma \propto \sqrt{G\mSMBH/a}$. Combining all of the above gives
\begin{align}
    T_J \propto (1 - \eEMRI)^{5/2 - \delta}(1 + \eEMRI)\mSMBH^{1.5} \rpEMRI^{\delta - 3/2} \,,
\end{align}
where we have used $\rpEMRI = \aEMRI(1 - \eEMRI)$, whence Eq.~\eqref{eq:T_GW} yields
\begin{align}
    \frac{T_J}{T_{\rm GW}} \propto \frac{\mSMBH^{3.5} (1-\eEMRI)^{3-\delta} \rpEMRI^{\delta - 11/2}}{(1 + \eEMRI)^{5/2}} = \frac{\mSMBH^{3.5} \aEMRI^{\delta - 3} \rpEMRI^{- 5/2}}{(1 + \eEMRI)^{5/2}}\,. \label{eq:T_J/T_GW}
\end{align}
While Eq.~\eqref{eq:T_GW} is only strictly valid in the limit $1-\eEMRI \ll 1$, it proves useful for our discussion as almost all EMRIs are in this regime when crossing the loss-cone boundary.

First of all, notice that the second relation in Eq.~\eqref{eq:T_J/T_GW} implies that a smaller EMRI periapsis corresponds to a larger EMRI semi-major axis for a given $\mSMBH$ and fixed $T_J/T_{\rm GW}=1$. This follows from the second relation in Eq.~\eqref{eq:T_J/T_GW} where the exponents of $\aEMRI$ and $\rpEMRI$ have the same sign (for $\delta < 3$, which almost any realistic stellar cusp obeys). Thus, as $\rpEMRI$ decreases, $\aEMRI$ (and hence eccentricity) must increase to maintain $T_J/T_{\rm GW} = 1$. It follows that EMRIs with small (large) initial periapsis are coming from the outer (inner) regions of the SoI and hence have larger (smaller) eccentricities. In fact, the right cut-off is a consequence of the condition $\eEMRI \geq 0$. This is all shown in Figure \ref{fig:f_rp_dists_a_e}.

\begin{figure}
    \centering
    \subcaptionbox{Distribution of EMRI semi-major axis, $\aEMRI_0$.\label{subfig:f_rp_dists_e}}
    [0.9\linewidth]
    {\includegraphics[width=\linewidth]{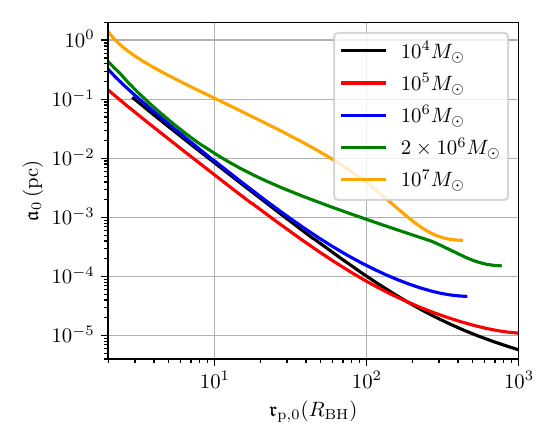}}

    \vspace{0.5em}

    \subcaptionbox{Distribution of EMRI eccentricity, $\eEMRIz = e_{\rm LC}$.\label{subfig:f_rp_dists_a}}
    [0.9\linewidth]
    {\includegraphics[width=\linewidth]{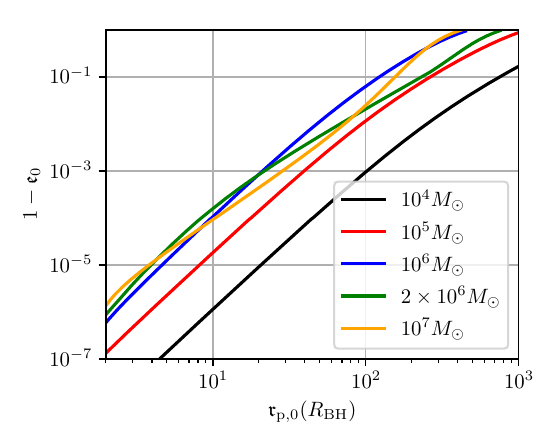}}

    \caption{The EMRI initial semi-major axis (a) and EMRI initial eccentricity (b) as a function of the initial periapsis for $\mSMBH = [10^4, 10^5, 10^6, 2 \times 10^6, 10^7]\,\Msolar$ at $t = 2.14 \, \Gyr$.}
    \label{fig:f_rp_dists_a_e}
\end{figure}

Further, the first relation in Eq.~\eqref{eq:T_J/T_GW} implies that, for a given eccentricity, $M_{\rm BH}^{\delta - 2} (\rpEMRI/\RSMBH)^{\delta - 11/2}$ must remain constant. Since the stellar component predominately responsible for inducing relaxation will exhibit a BW $\delta = 7/4$ profile when relaxed (at least in a two-mass stellar component; cf.~\S\ref{subsubsec:SSE}), then the exponents of $M_{\rm BH}$ and $\rpEMRI/\RSMBH$ will always have the same sign, meaning that an increase in $M_{\rm BH}$ will cause a decrease in $\rpEMRI/\RSMBH$, explaining why the right cut-off decreases as we move from $\mSMBH = 10^4 \, \Msolar$ up to $\mSMBH = 10^6 \, \Msolar$. However, this behaviour is violated once $\mSMBH$ reaches $\mUR = 1.7 \times 10^6 \, \Msolar$ as in that case the cluster is no longer relaxed by $t = 2.14 \, \Gyr$.

Due to relaxation occurring in the outer regions more quickly than the inner regions (see the left panels of Figure \ref{fig:DM_threepanel_plots_heavy} and Figure \ref{fig:DM_threepanel_plots_light}), for $\mSMBH \geq \mUR$, the initial condition is frozen in only in the inner region, while it is frozen-in at all radii for even larger $\mSMBH$. Thus, for $\mSMBH$ just above $\mUR$, $\langle \rho(a) \rangle$ drops for small $\aEMRI$ (larger $\rpEMRI$) only. This increases $T_J$ here, since $T_J \propto 1/\langle \rho(a) \rangle$, and hence increases the proportionality constant in Eq.~\eqref{eq:T_J/T_GW} which allows $\rpEMRI$ to reach larger values before $T_{\rm GW} = T_J$, explaining why the right cut-off suddenly jumps outwards between $\mSMBH = 10^6 \, \Msolar$ and $\mSMBH = 2 \times 10^6 \, \Msolar$ in Figure \ref{fig:f_rp_dists}. As we increase mass further, this right cut-off continues to move inwards for the same reason it did for relaxed systems (except with $\delta = 1$ due to our selection of a Hernquist initial profile).

Finally, we see that all EMRI distributions are dominated by smaller periapsides. This is due to a complex cocktail of effects: 1) the complicated parameter dependence of the sink term, $\nu(E)$, in Eq.~\eqref{eq:nu(E)}, 2) mass segregation causing heavier objects, such as the sBHs that make up EMRIs, to move inwards (if the system has had time to relax), and 3) the outer regions of the SoI often dominating the SoI's mass (once volume-weighted) despite the density profile scaling inversely with semi-major axis. However, the difference in relaxed and unrelaxed distributions can be explained quite simply. It is merely a consequence of mass segregation occurring more slowly within the inner regions of the SoI compared to the outer regions. As a result, as we move across the $\mUR$ boundary, we see a reduction in EMRIs with small semi-major axis (large periapsis) before we see a reduction in those with large semi-major axis (small periapsis). Thus, the EMRI distribution is completely dominated by small periapsides for $\mSMBH = 2 \times 10^6 \, \Msolar$. As $\mSMBH$ increases further, however, we find no mass segregation occurs at any semi-major axes, thus neutralising the relative deficit at large periapsides.

\section{Ejection Fraction Scaling with Mass Ratio}
\label{appsec:Scaling with Mass Ratio}

Here, we provide analytic arguments as to why we expect the ejection probability of a single DM particle over a single EMRI's orbit to be (explicitly) independent of $\mSMBH$ and to depend only on $q^2$, where $q = M_*/\mSMBH$ is the EMRI mass ratio. Knowing this relation allows us to extrapolate ejection fractions for arbitrary $q \ll 1$ by knowing it for a single $q \ll 1$.

The Schwarzschild metric is given by
\begin{align}
    \dd s^2 = - \left(1 - \frac{\RSMBH}{r}\right) \mathrm{c}^2 \dd t^2 + \frac{1}{1 - \frac{\RSMBH}{r}} & \dd r^2 + r^2 \dd \Omega^2 \,.
\end{align}
Upon applying the coordinate transformation $(t, r) \mapsto (\tilde{t}, \tilde{r}) \equiv (ct/\RSMBH, r/\RSMBH)$, we get 
\begin{align}
    \dd s^2 = \RSMBH^2 \left(- \left(1 - \frac{1}{\tilde{r}}\right) \dd \tilde{t}^2 + \frac{1}{1 - \frac{1}{\tilde{r}}} \dd \tilde{r}^2 + \tilde{r}^2 \dd \Omega^2\right) \,.
\end{align}
This metric implies that the Christoffel symbols, $\tilde{\Gamma}^\mu{ }_{\alpha \beta}$, are independent of $\RSMBH$, and hence so is the geodesic equation. 
It follows that the evolution of test particles in a Schwarzschild space-time is independent of the MBH mass, provided time and distance are measured in units of $\RSMBH/c$ and $\RSMBH$, respectively \footnote{More generally, this holds for any spacetime in general relativity described by a single dimensional scale, including the Kerr metric.}. However, the mass of the EMRI, or more precisely the mass ratio $q = \mEMRI/\mSMBH$, plays a crucial role in the MBH--sBH--DM evolution, since heavier sBHs are more likely to eject or swallow DM. 
To obtain an analytic scaling of the ejection probability of a single DM particle during a single EMRI orbit with the mass ratio $q$, it is useful to imagine a disc representing the DM particle's orbit after being averaged over radius and azimuth (due to relativistic apsidal precession), with the inner and outer radii given by $\rpDM$ and $\raDM$, respectively. This disc has an associated radial density, $\sigma_{\rm DM}(r)$, representing the probability density of finding the DM particle at a particular radius at any given time \citep{Kocsis_Tremaine2015}
\begin{align}
    \sigma_{\rm DM}(r) = \frac{r}{\pi a} \frac{1}{\sqrt{(r - \rpDM)(\raDM-r)}}\,,
\end{align}
where $a=(\rpDM+\raDM)/2$ is the semi-major axis of the DM particle. The probability density of finding an sBH particle at some radius $r$ is given similarly by $\sigma_{\rm EMRI}$ with $(\rpDM,\raDM) \to (\rpEMRI,\raEMRI)$.

Consider a close DM-sBH encounter and define $b_{\rm max}$ to be the maximum impact parameter that the DM particle can have with the sBH such that it receives a large enough velocity kick to be ejected from the system. Such an encounter requires the orbital radii $r$ of the DM and EMRI particles to be approximately within $b_{\rm max}$, and the azimuthal angles to be aligned to within $b_{\rm max}/(\pi r)$. Hence the instantaneous ejection probability per orbit is 
\begin{align}
    p_{\rm eject} &= 
    \int^{\min(\raEMRI,\raDM)}_{\max(\rpEMRI,\rpDM)} dr \,\frac{b_{\rm max}}{\pi r} \times \nonumber\\
    &\qquad \qquad \int_{0}^{b_{\max}} d\xi \;
    \sigma_{\rm DM}(r)\sigma_{\rm EMRI}(r+\xi)\,, \nonumber\\
    &\approx
    \int^{\min(\raEMRI,\raDM)}_{\max(\rpEMRI,\rpDM)} dr \,\frac{b_{\rm max}^2}{\pi r}
    \sigma_{\rm DM}(r)\sigma_{\rm EMRI}(r) + \mathcal{O}(b_{\rm max}^3)\,,
\end{align}
and $p_{\rm eject} = 0$ if $\max(\rpEMRI,\rpDM)>\min(\raEMRI,\raDM)$. This estimate assumes that the azimuthal angle is uniformly distributed, which is valid once averaging the orbit over the apsidal precession cycle. Thus, to leading order, $p_{\rm eject} \propto b_{\rm max}^2$.

In order to calculate $b_{\rm max}$, we use a Newtonian approach. The impulse kick imparted on a DM particle by a passing sBH with mass $\mEMRI$, relative velocity $V$ and impact parameter $b$ is \citep[][Eq.~(3.53)]{Binney_2008}
\begin{align}
    |\Delta V| = \frac{2 \mathrm{G} \mEMRI V}{\sqrt{\mathrm{G}^2 \mEMRI^2 + b^2 V^4}} = \frac{2V}{\sqrt{1 + b^2/b_{90}^2}}\,,
\end{align}
where $b_{90}= \mathrm{G}M_*/V^2$ is the impact parameter for a $90^{\circ}$ deflection. Ejection requires $|\Delta V| \sim V$, which implies the maximum impact parameter for ejection is $b_{\rm max} \sim b_{90}$. Since the sBH orbits the MBH, $V^2 \sim \mathrm{G}\mSMBH/\rEMRI$ where $\rEMRI$ is the sBH's distance from the MBH at the moment of the interaction, and hence
\begin{align}
    b_{\max} \sim q \rEMRI \,.
\end{align}
It follows that the ejection probability scales as $b_{\max}^2 \propto q^2$.

Finally, we expect the proportionality constant to increase as inclination decreases. This is because lower inclinations will (i) cause the sBH to penetrate a larger surface area of the DM orbital disc; (ii) allow for a larger range of longitudes of ascending nodes to undergo a strong interaction with the sBH; and (iii) cause a smaller relative velocity between the DM particles and the sBH. Each of these effects increases the interaction cross section but does not alter its fundamental dependence on $b_{\rm max}^2$. This is all numerically verified in \S\ref{subsubsec:Scaling with Mass Ratio}, namely through Figure \ref{fig:f_ej_scaling}.

\section{Time Evolution of an EMRI}
\label{appsec:Time Evolution of an EMRI}

\begin{figure}
    \centering
    \includegraphics[width=0.49\textwidth]{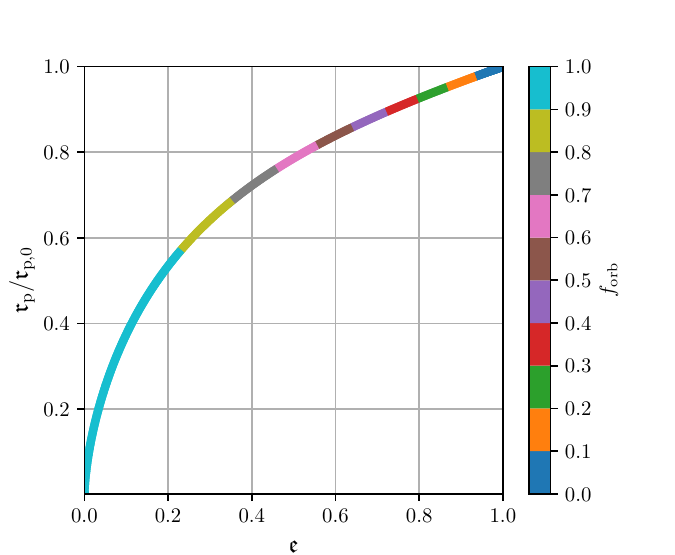}
     \caption{The evolution of an EMRI in $\eEMRI - \rpEMRI$ space. The colour corresponds to the fraction of total orbits completed, $f_{\rm orb} = N_{\rm orb}/N_{\rm total}$.}
    \label{fig:GW_rp_vs_e}
\end{figure}

Here, we derive two relations characterising the evolution of an EMRI undergoing GW-induced inspiral. First, we determine the evolution of the EMRI's periapsis and eccentricity as a function of the number of orbits completed, as required in Eq.~\eqref{eq:ln P_stay EMRI lifetime}. We find that, for a given initial eccentricity, the dimensionless evolution is independent of the component masses as well as the initial periapsis, and only dependent on the fraction of total orbits completed. Second, we find that $N_{\rm total}$ scales as $q^{-1}$ (Eq.~\ref{eqn:N total with q}).

To this end, we begin with the orbit-averaged rate of change of semi-major axis, $\aEMRI$, and eccentricity, $\eEMRI$, due to GW energy dissipation \citep{Peters1964}
\begin{align}
    \langle \dot{\aEMRI} \rangle &= - \frac{\beta}{\aEMRI^3 (1-\eEMRI^2)^{7/2}} \left[1 + \frac{73}{24}\eEMRI^2 + \frac{37}{96}\eEMRI^4\right], \label{eq:adot GW}\\
    \langle \dot{\eEMRI} \rangle &= - \frac{19}{12} \frac{\beta\eEMRI}{\aEMRI^4 (1-\eEMRI^2)^{5/2}} \left[1 + \frac{121}{304} \eEMRI^2\right]\,, \label{eq:edot GW}
\end{align}
where
\begin{align}
    \beta = \frac{64}{5} \frac{\mathrm{G}^3 \mSMBH \mEMRI(\mSMBH + \mEMRI)}{\mathrm{c}^5}\,.
\end{align}
Furthermore, these equations imply that \citep{Peters1964}
\begin{align}
    \frac{\aEMRI(\eEMRI)}{\aEMRI_0} = \frac{k(\eEMRI)}{k(\eEMRIz)}
\end{align}
where
\begin{align}
    k(\eEMRI) = \frac{\eEMRI^{12/19}}{1-\eEMRI^2}\left(1+\frac{121}{304}\eEMRI^2\right)^{870/2299}
\end{align}
where $\aEMRI_0$ and $\eEMRIz$ are the EMRI's initial semi-major axis and eccentricity, respectively. Clearly, for given $\eEMRIz$, $\tilde{\aEMRI}(\eEMRI) = \aEMRI(\eEMRI)/\aEMRI_0$ is independent of the component masses and $\aEMRI_0$. 

Using Kepler's law (which becomes increasingly inaccurate as orbits become more relativistic), the rate of change of the number of orbits completed, $N_{\rm orb}$, is 
\begin{align}
    \frac{\dd N_{\rm orb}}{\dd t} = \frac{n}{2\pi} = \frac{1}{2\pi}\sqrt{\frac{\mathrm{G}(\mSMBH + \mEMRI)}{\aEMRI^3}} \,, \label{eq:dN_orb/dt}
\end{align}
where $n$ is the mean motion frequency. Thus the accumulated number of orbits from an initial $\eEMRIz$ to any $\eEMRI$ is 
\begin{align}
    N_{\rm orb}(\eEMRI) &= \int^\eEMRI_{\eEMRIz} \left<\frac{\dd N_{\rm orb}}{\dd t}\right> \frac{1}{\langle\dot e \rangle} \dd e \\
    &\approx 2.81 \times 10^6 \left(\frac{\aEMRI_0/\RSMBH}{100}\right)^{\frac{5}{2}} \left(\frac{10^{-3}}{q}\right) K(\eEMRI)
\end{align}
for $q \ll 1$ where
\begin{align}
    K(\eEMRI) = \int_\eEMRI^{\eEMRIz}\left(\frac{k(e)}{k(\eEMRIz)}\right)^{5/2}
    \frac{ (1-e^2)^{5/2}}{ e(1 + \frac{121}{304} e^2)} \dd e \,.
\end{align}
The fraction of orbits completed is then $f_{\rm orb} = K(\eEMRI)/K(0)$, which is independent of the component masses and $\aEMRI_0$. Finally, $f_{\rm orb}$ is invertible since $K(\eEMRI)$ is monotonic, and thus the evolution of $\eEMRI$ only depends on $f_{\rm orb}$ and $\eEMRIz$ and, as a result, so does $\tilde{\aEMRI}$ and $\rpEMRI/\rpEMRIz$, where $\rpEMRIz$ is the initial EMRI periapsis. This simple parameter dependence allows for quick computation of EMRI eccentricity and periapsis evolution. 

Figure~\ref{fig:GW_rp_vs_e} shows the evolution of an initially parabolic EMRI in $\eEMRI - \rpEMRI$ space, where $\rpEMRIz$ is the initial periapsis, through direct numerical integration of Eqs.~\eqref{eq:adot GW}, \eqref{eq:edot GW}, and \eqref{eq:dN_orb/dt}. As an aside, we can see that highly eccentric orbits circularise rapidly. Consequently, the apoapsis of eccentric orbits decays rapidly compared to the periapsis.

Finally, the total number of orbits completed is $N_{\rm total} = N_{\rm orb}(\eEMRI = 0)$. For a fixed MBH mass $\mSMBH$, this is proportional to $1/(q\sqrt{1+q})$, which scales as $1/q$ for $q \ll 1$. 

\section{Validating Orbit-Independent Ejection Probabilities}
\label{appsec:Validating Orbit-Independent Ejection Probabilities}

While this manuscript was on the arXiv and under review, \citet{Karydas_2026} appeared on the arXiv. They consider the same physical problem but model DM depletion using {\tt HaloFeedback}, a secular code that evolves the DM distribution during an EMRI's inspiral via a Fokker-Planck-type approach that includes both strong and weak interactions. They claim to have found results that differ from ours and that this difference is attributed to our assumption of independent ejection probabilities per orbit. However, we compare our results to full-inspiral $N$-body simulations and find good agreement, justifying our assumption. We also compare predictions for the depletion around a Sgr A$^*$-like MBH after 270 total EMRIs. Our work predicts slightly larger density reductions, but this can most likely be attributed to the different EMRI distributions that are used. Overall, we do not find disagreement with their results or with direct full-inspiral $N$-body simulations.

We first validate the assumption of independent orbit ejection probabilities --- specifically in Eq.~\eqref{eq:ln P_stay EMRI lifetime} --- by comparing our single-orbit calculation of $- \ln(\langle P_{\rm stay} \rangle_{\iota})/N_{\rm orbs} = q^2 \langle A \rangle_{{\rm orbs, \, eff}}$ in Eq.~\eqref{eq:A_orb,eff} to that found using direct full-inspiral $N$-body simulations, where $\langle P_{\rm stay} \rangle_{\iota}$ is the inclination-averaged fraction of DM particles that remain. In Figure \ref{fig:remov_lnf_N_2}, we show this value calculated using $2.3 \times 10^4$ full-inspiral $N$-body simulations (blue) against the predicted value given by our model (red) for $q = 10^{-3}$, $\eDM = 0.2$, $\rpDM = 50 \, \RSMBH$, $\eEMRI = 0$, and $\rpEMRI = 75 \, \RSMBH$. There is an initial disagreement for $N_{\rm orbits} \lesssim 10^3$. This appears because the finite number of particles used makes ejection fractions incredibly noisy for small $N_{\rm orbits}$. There is also an additional effect that results from us having randomised over the argument of periapsis and longitude of the ascending node when calculating $A$, which is only valid over time-scales much longer than the apsidal advance time-scale, which is $\sim 30$ EMRI orbits. However, its presence is overwhelmed by the noise here. Above $10^3$ orbits, where we are no longer susceptible to noise, we find strong agreement. We note that any EMRI, even with a large mass ratio of $q = 10^{-3}$ and zero eccentricity, will exhibit more than 1000 orbits if its initial periapsis is greater than $\sim 4.3\RSMBH$. This justifies the assumption of independent orbits.

\citet{Karydas_2026} consider a single system with an MBH mass of $\mSMBH = 4 \times 10^6 \, \Msolar$ and an EMRI rate of $27\,\Gyr^{-1}$ for $10 \, \Gyr$, giving a \textit{total} of 270 EMRIs. As stated above, most predicted EMRI rates sit around (100-300)$\,\Gyr^{-1}$ (c.f. \S\ref{subsec:Param Distributions}). They obtain a maximum density reduction of approximately $18\%$. Using the EMRI distribution predicted by our FP model and the same number of total EMRIs, we find a slightly greater reduction of approximately $50\%$, as shown in Fig.~\ref{fig:Karydas_2026_z} --- an $\mathcal{O}(\mbox{a few})$ disagreement at most. However, this can be attributed to the different EMRI distributions used. \citet{Karydas_2026} draw the EMRI eccentricities from ref.~\cite{Hopman_2005} at $a=8.5\,\RSMBH$ and evolve them backwards using Eqs.~\eqref{eq:adot GW} and \eqref{eq:edot GW} to $a=360\,\RSMBH$, whereas we use our FP model. Furthermore, at $\mSMBH=4\times10^6\,\Msolar$ our nuclear star cluster has not relaxed by $2.14\,\Gyr$, so its predicted EMRI distribution remains uncertain and sensitive to the assumed initial conditions, as discussed in Appendix~\ref{appsec:Diff NSC ICs}. It would be interesting to compare the predicted densities over a wider range of MBH masses, such as $10^4\,\Msolar$--$10^7\,\Msolar$, and EMRI rates, such as $1\,\Gyr^{-1}$--$1000\,\Gyr^{-1}$, which we have readily available, as shown in \S\ref{sec:DM and EMRIs}.

\begin{figure}[h!]
    \centering
    {\includegraphics[width=0.48\textwidth]{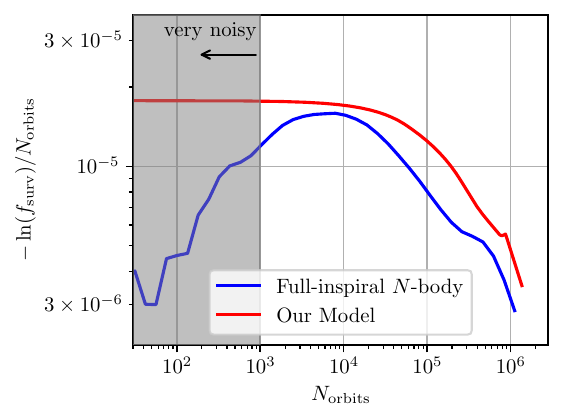}}
    \caption{
    The evolution of $-\log f_{\rm surv}/N_{\rm orbits}$ as a function of the number of orbits exhibited by the EMRI, $N_{\rm orbits}$. The blue curve, pertaining to the numerical result, was obtained by integrating $2.3 \times 10^4$ full-inspiral $N$-body simulations using \multistar, for $q = 10^{-3}$, $\eDM = 0.2$, $\rpDM = 50 \, \RSMBH$, $\eEMRI = 0$, and $\rpEMRI = 75 \, \RSMBH$. The red curve shows the prediction of our model for the same parameters. The region below $N_{\rm orbits} = 10^3$ is shaded grey because of noise susceptibility, however all EMRIs are expected to exhibit many more than $10^3$ orbits (see text).
    }
    \label{fig:remov_lnf_N_2}
\end{figure}

\begin{figure}[h!]
    \centering
    {\includegraphics[width=0.49\textwidth]{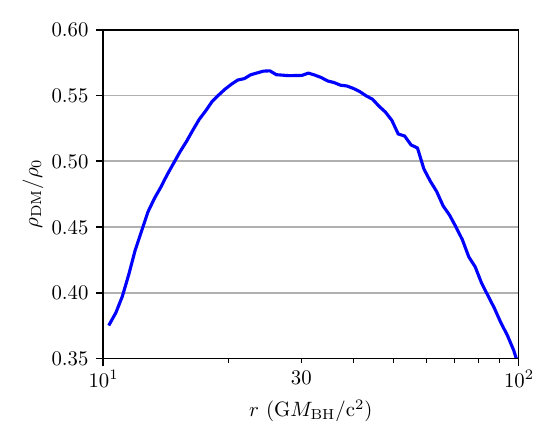}}
    \caption{
    The fractional reduction in DM density as a function of radius for
    $\mSMBH=4\times10^6\,\Msolar$ after 270 EMRIs. Our calculation predicts
    a density reduction of $\sim 50\%$ over the radii shown, whereas
    \citet{Karydas_2026} report a maximum reduction of approximately $18\%$. This discrepancy is most likely attributable to the use of different distributions for EMRI periapsides and eccentricities.
    }
    \label{fig:Karydas_2026_z}
\end{figure}

\end{document}